\documentclass{aa}  

\usepackage{graphicx}
\usepackage{txfonts}
\usepackage{lipsum}
\usepackage{subcaption}         
\usepackage{lscape}             
\usepackage{placeins}           
                               
\begin{document}

\let\linenumbers\relax

   \title{Enhancing Photometric Redshift Estimation for LSST with a Hybrid LSTM-Mixture Density Network}


%
%
%

   \author{Zhijian Luo\inst{1}\corrauth{zjluo@shnu.edu.cn}        
        \and Yangyang Li\inst{1}\email{liyangyang11223@163.com} %
        \and Xinyu Luo\inst{1}\email{2313047862@qq.com}
        \and Hubing Xiao\inst{1}\corrauth{hubing.xiao@outlook.com}
        \and Wei Fang\inst{1,2}\email{cat@nestor-edp.org}
        \and Wenxiang Pei\inst{1}\corrauth{wxpei@nao.cas.cn}
        \and Shaohua Zhang\inst{1}\email{zhangshaohua@shnu.edu.cn}
        \and Chenggang Shu \inst{1}\email{cgshu@shao.ac.cn}
        }

   \institute{Shanghai Key Lab for Astrophysics, Shanghai Normal University, 200234, Shanghai, People’s Republic of China
   \and Department of Physics, Shanghai Normal University, 200234, Shanghai, People’s Republic of China}

 
  \abstract
   {Accurate photometric redshift (photo-$z$) estimation and robust uncertainty quantification are essential for the Legacy Survey of Space and Time (LSST) to achieve its precision cosmology goals. Traditional machine learning algorithms are largely restricted to point estimates, struggling to characterize the multimodal nature of redshift probability density functions (PDFs) and the degeneracies within the color–redshift space.
   }
   {We validate the LSTM-MDNz architecture, which integrates sequential feature extraction with flexible probability density modeling. The goal is to enhance both prediction accuracy and uncertainty calibration across a broad redshift range, meeting the stringent data quality requirements necessitated by next-generation cosmological analysis.} 
   {The LSTM-MDNz framework treats multi-band photometry as wavelength-ordered sequences, utilizing long short-term memory (LSTM) networks to capture non-linear evolutionary correlations across the spectral energy distribution (SED). A mixture density network (MDN) is then employed to explicitly model posterior PDFs via Gaussian mixture models (GMMs). Performance is evaluated on the Hyper Suprime-Cam (HSC) GalaxiesML dataset— which server as a small-scale proxy for next-generation surveys like LSST—and benchmarked against the Bayesian neural network (BNN) architecture established by \citet{2024ApJ...964..130J}.}
   {The proposed model consistently outperforms the BNN baseline, achieving a $\sim$10\% improvement in point-estimation accuracy (specifically across RMSE, MAE, scatter, and $\sigma_{\rm NMAD}$) and a $\sim$20\% reduction in both outlier and catastrophic outlier rates. A uniform probability integral transform (PIT) distribution confirms well-calibrated probabilistic outputs. Furthermore, the PDF-based confidence metric $z_{\rm conf}$ enables high-purity catalog construction: excluding just approximately 4\% of extremely low-confidence ($z_{\rm conf} < 0.05$) samples reduces the overall outlier rate by $\sim$48\%.}
   {LSTM-MDNz offers a scalable and high-precision approach for photometric redshift estimation in LSST-scale surveys. Its robust uncertainty quantification capability provides important support for precision cosmology, particularly in advancing weak lensing and large-scale structure studies.}

   \maketitle

\section{Introduction}
The precise measurement of galaxy redshifts is a cornerstone of modern observational cosmology. With the onset of next-generation large-scale survey projects—including the European Space Agency’s (ESA) Euclid mission \citep{2011arXiv1110.3193L,2025A&A...697A...1E}, the Vera C. Rubin Observatory’s LSST \citep{2019ApJ...873..111I}, the Chinese Space Station Telescope (CSST; \citealt{zhan2011consideration,2018MNRAS.480.2178C,2026SCPMA..6939501C}), and the Nancy Grace Roman Space Telescope \citep{2019arXiv190205569A}—the astronomical community is poised to acquire multi-band photometric data for billions of galaxies. While spectroscopic observations provide the most reliable redshifts, their prohibitive observational cost renders obtaining spectra for such vast numbers of galaxies impractical. Consequently, photometric redshift (photo-$z$) techniques—which indirectly estimate redshifts from photometric features such as magnitudes, colors, and morphology—have become indispensable for investigating galaxy evolution, probing the nature of dark energy, and constraining cosmological parameters through weak gravitational lensing and large-scale structure measurements \citep{2022ARA&A..60..363N}.

However, photometric redshift ($z_{\rm phot}$) estimation faces inherent challenges in both precision and reliability. Because photometric surveys rely on a limited set of broad-band filters to discretely sample a galaxy's spectral energy distribution (SED), this low-resolution representation of critical features—such as the 4000 Å break—inevitably introduces color–redshift degeneracies. These degeneracies result in multi-modal probability distributions where distinct redshift solutions yield identical observed colors, leading to systematic biases (sometimes subtle) as well as extreme predictive failures. To rigorously characterize these uncertainties, a multi-dimensional metric system is typically employed: normalized median absolute deviation ($\sigma_{\rm NMAD}$) to quantify statistical scatter, median bias ($\langle \Delta z \rangle$) to capture systematic offsets, and the outlier fraction ($O$) to track the frequency of significant predictive failures, among others.

To keep these errors within the limits required for cosmological analysis, projects such as LSST have established stringent requirements in their science requirements documents (SRDs). For a galaxy sample with an $i$-band limiting magnitude of $i < 25.3$ over the redshift range $0 < z < 3$, LSST requires \citep{2018AJ....155....1G,2020AJ....159..258G,2026AJ....171..114S}: (1) a median bias of $|\Delta z| < 0.003$, where $\Delta z = (z_{\rm pred} - z_{\rm true})/(1 + z_{\rm true})$; (2) a scaled photo-$z$ scatter of $\sigma_{\Delta z} < 0.02$ \citep{2009arXiv0912.0201L}; and (3) a $3\sigma$ catastrophic outlier rate below 10\% \citep{2006MNRAS.366..101H,2006ApJ...636...21M,2007MNRAS.381.1018A,2008MNRAS.390..149K}. Similarly, the Euclid mission imposes an even more stringent median bias requirement of $|\Delta z| < 0.002$ within the range $0.2 < z < 2.5$ \citep{2017MNRAS.468.4323B,2025A&A...697A...1E}. These demanding standards pose a significant challenge to the robustness and accuracy of current photometric redshift algorithms.

Currently, photometric redshift estimation methods are broadly categorized into two main classes: template fitting and machine learning. Template fitting methods—such as HyperZ \citep{2000A&A...363..476B}, EAZY \citep{2008ApJ...686.1503B}, Mizuki \citep{2020arXiv200301511N}, BPZ \citep{2000ApJ...536..571B}, ZEBRA \citep{2006MNRAS.372..565F}, and LePhare \citep{1999MNRAS.310..540A}—fit observed multi-band photometric data against a library of SED templates, typically using $\chi^2$ minimization to treat redshift as a free parameter. These methods offer clear physical interpretability and do not require large spectroscopic training samples, making them well suited for deep-field surveys or studies of rare objects. However, their accuracy is significantly limited by the completeness and precision of the template libraries, as well as by the simplified treatment of physical processes such as intrinsic galactic extinction.

In contrast, machine learning-based approaches—including algorithms such as TPZ \citep{2013MNRAS.432.1483C}, ANNz2 \citep{2016PASP..128j4502S}, SVM \citep{2017A&A...600A.113J}, DEmP \citep{2018PASJ...70S...9T}, Random Forests \citep{2001MachL..45....5B, 2013MNRAS.432.1483C, 2015MNRAS.452.3710R, 2021MNRAS.502.2770M, 2024MNRAS.52712140L}, and more recent deep learning architectures such as Delight \citep{2017ApJ...838....5L}, CNNs \citep{2015MNRAS.452.4183H, 2018A&A...609A.111D, 2019A&A...621A..26P, 2021ApJ...909...53Z}, BNNs \citep{zhou2022photometricBNN}, and LSTMs \citep{2024MNRAS.535.1844L,2026ApJS..282...46C}—construct complex, non-linear frameworks to directly learn the mapping between multi-dimensional photometric attributes (e.g., magnitudes, colors, and morphology) and redshifts. By leveraging large-scale spectroscopic samples as training data, these models bypass explicit physical modeling. Provided that the training samples are sufficiently comprehensive and representative of the target population, machine learning methods generally achieve higher predictive accuracy and greater statistical robustness than traditional template-based techniques.

As the precision requirements for cosmological analyses continue to increase, relying solely on redshift point estimates is no longer sufficient to meet the demands of next-generation surveys. The redshift probability density function (PDF) provides a more comprehensive characterization of the posterior distribution, effectively capturing uncertainties arising from both statistical noise and physical degeneracies—such as the Lyman–Balmer break ambiguity. In precision cosmological studies, including weak gravitational lensing and large-scale galaxy clustering, utilizing the full PDF enables a more accurate marginalization over redshift uncertainties. This approach effectively mitigates the propagation of redshift errors into cosmological parameter estimation, thereby significantly enhancing the fidelity of the resulting measurements \citep{2018ARA&A..56..393M, 2015MNRAS.449.1043B}.

To simultaneously achieve high-precision point estimation and reliable PDF characterization, probabilistic deep learning frameworks have gained significant attention in the field of photometric redshift estimation. Among these, Bayesian neural networks (BNNs) provide a rigorous statistical foundation for capturing epistemic uncertainty by treating network weights as probability distributions rather than deterministic values. For instance, \citet{2024ApJ...964..130J} utilized Hyper Suprime-Cam (HSC) data as a precursor sample for LSST, demonstrating the robustness of BNNs in addressing the challenges posed by future large-scale survey missions. Expanding on this, \citet{2024arXiv241118054S, 2026AJ....171..114S} investigated multi-source training strategies by combining spectroscopic data from GalaxiesML \citep{2024arXiv241000271D} with high-precision photometric redshifts (TransferZ) from COSMOS2020. Their findings demonstrate that composite dataset training effectively mitigates spectroscopic representativeness bias, significantly outperforming single-source training or transfer learning within $0.3 < z < 1.5$. Additionally, they utilized split conformal prediction to calibrate uncertainty estimates, ensuring statistically reliable intervals for both BNNs and deterministic models.

Despite these advancements, BNN-based models still face certain limitations in practical applications. First, regarding feature extraction, traditional BNN architectures typically rely on fully connected layers, which treat multi-band photometric measurements as discrete and independent input features. This approach is often suboptimal for incorporating observational uncertainties and fails to fully account for the intrinsic physical correlations arising from the continuous spectral evolution across different filters. By neglecting the sequential nature of wavelength-ordered data, such architectures may struggle to capture subtle variations in SEDs. These limitations can weaken the model's ability to recognize complex patterns, ultimately compromising the overall precision and robustness of the resulting redshift estimates.

Second, regarding uncertainty quantification, many BNN implementations rely on simplified parametric assumptions. For instance, the architectures explored by \citet{2024ApJ...964..130J} typically characterize uncertainty through a single Gaussian posterior (defined by a mean and standard deviation). However, for LSST-like data, redshift PDFs are frequently non-Gaussian—often exhibiting heavy tails, skewness, or multimodality—particularly for faint sources where color–redshift degeneracies are most pronounced \citep{2013MNRAS.431.2766S}. Such unimodal simplifications may struggle to capture these complex degeneracies, potentially leading to elevated catastrophic outlier rates and compromised calibration in probability integral transform (PIT) distributions. Accurately modeling the multimodality of PDFs remains a central objective in meeting the precision metrics defined in the LSST SRD, especially for the faint galaxy populations that dominate deep-drilling fields.

To address these limitations, this study evaluates a novel hybrid deep learning architecture, LSTM-MDNz \citep{2026ApJS..282...46C}. Unlike traditional models with static inputs, this framework treats multi-band photometric fluxes and their associated observational uncertainties as wavelength-ordered sequences. By leveraging long short-term memory (LSTM) networks, the model effectively captures the nonlinear evolutionary trends inherent in SEDs. This sequential approach enables the extraction of deep representations that are implicitly constrained by the physical continuity of stellar populations across different frequency bands. 

On the inference side, the model integrates a mixture density network (MDN) to explicitly characterize the posterior redshift PDF. In contrast to the unimodal Gaussian assumptions prevalent in many BNN implementations, the MDN backend employs a Gaussian mixture model (GMM) to directly resolve the multimodal degeneracies and heavy-tailed distributions inherent in the data. Consequently, the LSTM-MDNz framework provides a robust characterization of physical uncertainties, balancing point-estimation precision with statistical reliability. This architecture offers a scalable analytical solution for the massive datasets anticipated from the LSST, providing the high-fidelity PDF information essential for achieving the survey's core cosmological goals.

To empirically validate the advantages of this architecture, we conduct a systematic evaluation of the LSTM-MDNz model using the HSC GalaxiesML dataset. This dataset serves as a high-fidelity, small-scale proxy for the photometric depth and filter configurations characteristic of the upcoming LSST. To ensure a rigorous and controlled benchmark, we strictly adhere to the experimental framework established by \citet{2024ApJ...964..130J}, employing an identical sample-splitting strategy to evaluate LSTM-MDNz against the BNN baseline under equivalent observational configurations. This approach ensures that any observed performance gains can be attributed to architectural and methodological improvements rather than variations in data pre-processing.

It is noteworthy that while the BNN baseline relies primarily on multi-band composite model (cmodel) magnitudes, the LSTM-MDNz architecture enables the explicit integration of cmodel magenitude error as supplementary input features. Since these uncertainty estimates are direct products of the photometric pipeline and require no additional observational overhead, this design allows the model to adaptively weight each band’s contribution based on its statistical signal-to-noise ratio (S/N). Consequently, LSTM-MDNz can effectively account for the heteroscedastic nature of the HSC data. Through this controlled comparison, we aim to quantify the performance gains achieved by maximizing the utility of existing photometric metadata and to demonstrate the superior reliability of the LSTM-MDNz framework within complex color spaces.

Our evaluation suite comprises a multi-dimensional set of metrics designed to assess both the precision of point estimates and the statistical reliability of the predicted PDFs. Point-estimation performance is quantified through standard indicators, including root mean square error (RMSE), mean absolute error (MAE), normalized median absolute deviation ($\sigma_{\rm NMAD}$), and the catastrophic outlier rate ($O_{\rm C}$), among others. To evaluate probabilistic calibration, we utilize the PIT distribution, while the continuous ranked probability score (CRPS) is employed to assess the simultaneous sharpness and accuracy of the uncertainty estimates.

To directly address the stringent outlier mitigation requirements defined in the LSST SRD, we propose a PDF-based confidence metric, $z_{\rm conf}$. This metric is specifically designed to isolate samples with high predictive ambiguity—particularly those characterized by multimodality or significant dispersion in their posterior distributions. By applying an optimized selection threshold, $z_{\rm conf}$ enables the construction of high-purity galaxy catalogs, effectively suppressing catastrophic outliers while maintaining a high degree of sample completeness. This filtering strategy provides a robust technical pathway toward meeting the cosmological precision requirements of next-generation wide-field surveys.

The structure of this paper is organized as follows: Section \ref{sec:data} describes the GalaxiesML dataset and the data preprocessing workflow; section \ref{sec:method} details the LSTM-MDNz model architecture and training strategy; Section \ref{sec:results} presents the results for both point estimation and probability density prediction, including a comparative analysis with the baseline model proposed by Jones et al. (2024). Within this section, we further introduce the PDF-based confidence metric $z_{\rm conf}$ and discuss its application in identifying high-risk samples and enhancing catalog purity. Finally, Section \ref{sec:conclusions} provides our conclusions.

\section{Data and Preprocessing} \label{sec:data}
\subsection{The GalaxiesML Dataset and Sample Selection}

In this study, we utilize the GalaxiesML dataset \citep{2024arXiv241000271D} as our primary experimental sample. This dataset is derived from the Hyper Suprime-Cam Public Data Release 2 (HSC-PDR2), which covers over 300 deg$^2$ with an $i$-band limiting magnitude of $\sim$26. Given that the multi-band configuration ($g, r, i, z, y$) and photometric depth of HSC-PDR2 are highly representative of the upcoming LSST, GalaxiesML offers one of the most substantial galaxy samples currently available for simulating LSST-like conditions. Consequently, it has been established as a critical precursor benchmark for evaluating machine learning algorithms tailored for next-generation large-scale surveys \citep{2024ApJ...964..130J, 2024arXiv241118054S, 2026AJ....171..114S}.

GalaxiesML is specifically optimized for deep learning applications under rigorous physical constraints. The sample was constructed by performing positional cross-matching (with a matching radius of $0.5''$) between HSC-PDR2 wide-field photometry and multiple spectroscopic redshift surveys. This integration provides spectroscopic ground truths from over ten authoritative surveys, including zCOSMOS \citep{2009ApJS..184..218L}, UDSz \citep{2013MNRAS.433..194B,2013MNRAS.428.1088M}, 3D-HST \citep{2014ApJS..214...24S,2016ApJS..225...27M}, FMOS-COSMOS \citep{2015ApJS..220...12S}, VVDS \citep{2013A&A...559A..14L}, VIPERS PDR1 \citep{2014A&A...562A..23G}, the Sloan Digital Sky Survey (SDSS) DR12/DR14Q \citep{2015ApJS..219...12A,2018A&A...613A..51P}, GAMA DR2 \citep{2015MNRAS.452.2087L}, WiggleZ DR1 \citep{2010MNRAS.401.1429D}, DEEP2 DR4 \citep{2003SPIE.4834..161D,2013ApJS..208....5N}, DEEP3 \citep{2011ApJS..193...14C,2012MNRAS.419.3018C}, and PRIMUS \citep{2011ApJ...741....8C,2013ApJ...767..118C}. To ensure the robustness and physical consistency of the dataset, \citet{2024arXiv241000271D} implemented stringent quality control measures, including the removal of unphysical outliers, deduplication of sources, and preservation of realistic signal-to-noise ratio (S/N) distributions. The resulting catalog contains 286,401 galaxies with high-quality spectroscopic redshifts (spec‑$z$), ensuring both reliable training labels and the statistical representativeness required for complex model architectures.

The dataset consists of two primary components: tabular data and image data, together forming a multimodal repository that integrates low-dimensional physical features with high-dimensional spatial morphologies. The tabular data, stored in CSV format, record galaxy identifiers and metadata alongside core photometric features—specifically $g, r, i, z, y$ cmodel magnitude measurements, their associated cmodel magnitude uncertainties, and pre-extracted morphological parameters. This format provides a versatile foundation for traditional statistical analysis and the deployment of feature-driven machine learning models, ranging from multi-layer perceptrons (MLPs) to gradient boosting decision trees (GBDTs).

Complementing the tabular records, the image data are encapsulated in HDF5 format, specifically optimized for high-throughput deep learning workflows. Beyond mirroring the scalar attributes, these HDF5 files provide five-band image tensors with spatial resolutions of either $127 \times 127$ or $64 \times 64$ pixels, preserving the intricate structural information of each galaxy. This dual-modality structure enables both the independent analysis of scalar properties and the implementation of sophisticated architectures that fuse pixel-level spatial information with photometrically derived feature vectors, facilitating a more comprehensive characterization of galaxy evolution.

   \begin{figure}[ht!]
   \centering
   \includegraphics[width=\hsize]{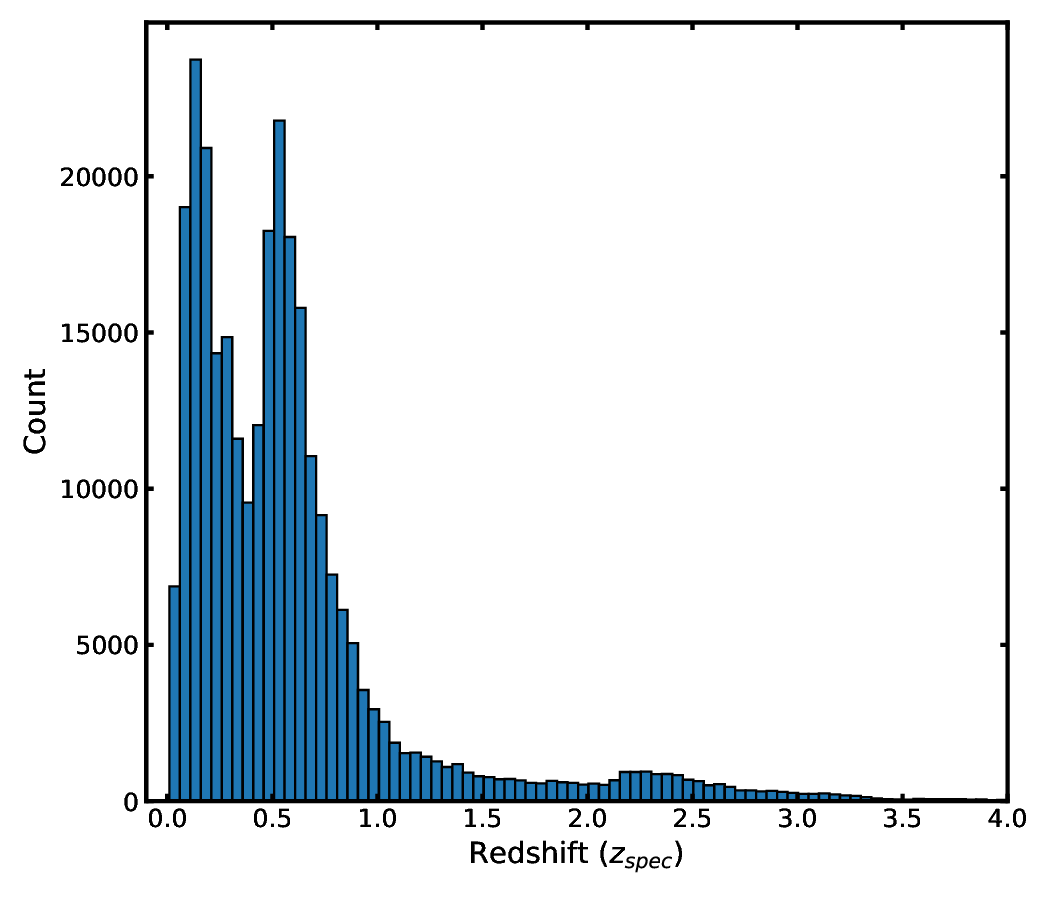}
      \caption{Redshift distribution of the GalaxiesML sample. The vertical axis represents the number of sources, and the horizontal axis represents the spectroscopic redshift ($z_{\rm spec}$).}
         \label{fig:z_dis}
   \end{figure}

The galaxies in the GalaxiesML dataset span a broad redshift range ($0.01 < z < 4.0$), with a statistical distribution exhibiting a prominent bimodal characteristic. As illustrated in Figure \ref{fig:z_dis}, the distribution features primary peaks near $z \approx 0.1$ and $z \approx 0.6$. This bimodal structure primarily arises from the diverse sample composition: the low-redshift peak is largely driven by the main galaxy sample (MGS) in the local universe, while the second peak near $z \approx 0.6$ corresponds to the high-density regimes of luminous red galaxies (LRGs) and emission-line galaxies (ELGs) \citep{2017MNRAS.468.4323B, 2026ApJS..282....6K}. Although the majority of galaxies are concentrated at $z < 1.0$, the density declines gradually beyond the second peak, becoming significantly sparse in the high-redshift regime ($z > 2.5$). This wide redshift distribution not only reflects the high-fidelity representation of the local universe but also includes a non-negligible high-redshift population. This enables the model to learn complex color–redshift mappings over an expansive look-back time, providing a rigorous testbed for evaluating model robustness under the significant sample imbalance inherent in deep-sky surveys.

It should be noted, however, that because this compiled spectroscopic sample integrates diverse historical surveys, its distribution reflects standard observational selection effects and localized completeness boundaries. The primary selection variations include a bright-end magnitude weighting ($i \lesssim 23.0$) typically required for reliable spectroscopic line identification, geometric targeting cuts (such as the intentional over-representation of LRGs and ELGs mentioned above), and a well-documented reduction in spectroscopic success rates within the intermediate redshift range ($1.2 < z < 2.0$). Compared to deeper, purely photometric reference surveys like the full HSC sample, this catalog exhibits lower completeness at the faint end ($i > 23.5$), where the multidimensional color space becomes sparsely sampled and largely bounded by the targeting footprints of the constituent surveys. In this study, we utilize this raw, uncorrected distribution directly to evaluate the baseline resilience and uncertainty quantification capabilities of the LSTM-MDNz framework when faced with realistic spectroscopic completeness boundaries.

In terms of data utilization, although GalaxiesML supports dual-modality inputs consisting of both images and photometry, this study specifically aims to explore the potential of sequence modeling for processing purely photometric data. Therefore, we focus on leveraging the multi-band photometric sequences and their associated uncertainties from the tabular data for redshift estimation. The integration of high-dimensional morphological features via multimodal deep learning will be pursued in future work.

\subsection{Data Preprocessing}

Prior to model training, we performed rigorous feature selection, sequential reconstruction, and numerical normalization. We selected the photometric cmodel magnitudes and their associated cmodel magnitude uncertainties for the five HSC optical bands ($g, r, i, z, y$) as primary features. While broader spectral coverage—such as ultraviolet data from GALEX or mid-infrared data from WISE—could further mitigate color–redshift degeneracies, we restricted our analysis to these five bands to ensure consistency with the expected depth and filter configurations of the future LSST survey \citep{2019ApJ...873..111I}. This maintains the same observational inputs as the BNN baseline \citep{2024ApJ...964..130J}, allowing for a controlled comparison of architectural efficiency.

Departing from the traditional treatment of multi-band features as isolated variables, this study adopts a physically inspired sequential design \citep{2024MNRAS.535.1844L, 2026ApJS..282...46C}. We arranged the $g, r, i, z, y$ data in increasing order of wavelength to construct a feature sequence of length five, where each timestep is represented by a two-dimensional vector, $\mathbf{v}_\lambda = [mag_{\lambda}, \Delta mag_{\lambda}]$. By explicitly coupling each magnitude with its associated measurement error, this sequential representation is designed to mimic the non-linear evolution of a galaxy's SED. Using the gating mechanisms of LSTM units, the model can adaptively weigh each band’s contribution based on its statistical signal-to-noise ratio (S/N) and capture inter-band gradients. This approach enables the MDN component to extract representations deeply rooted in stellar population physics, ultimately producing better-calibrated posterior PDFs within equivalent observational resource constraints.

Finally, to account for the inherent numerical distribution variances across different bands and features, we applied Z-score normalization to optimize training efficiency and numerical stability. The transformation is defined as:

\begin{equation}
     x' = \frac{x - \mu}{\sigma},
     \label{eq:normal} 
\end{equation}
where $\mu$ and $\sigma$ represent the mean and standard deviation of each feature, respectively. By scaling the feature space to a distribution with zero mean and unit variance, this preprocessing step facilitates more efficient model convergence. Furthermore, it promotes architectural robustness when processing samples with varying S/N, thereby supporting the consistency of both point-estimation accuracy and uncertainty calibration.

It is worth noting that since the mean and standard deviation are computed independently for each individual band, this transformation does not strictly preserve the raw physical shape of the galaxy SED along the wavelength axis. Mathematically, however, this independent $Z$-score normalization represents a deterministic, bijective mapping that preserves the underlying statistical cross-correlations and the entire information content across the filter set. The sequential network can successfully learn the continuity of the SED gradients within this transformed feature space, and the original physical profiles can be exactly reconstructed via inverse-normalization if required.

Following preprocessing, the GalaxiesML dataset (286,401 sources) was partitioned into training (80\%), validation (10\%), and test (10\%) subsets. To ensure a rigorous and unbiased benchmark, we strictly adopted the data partitioning protocol of \citet{2024ApJ...964..130J}, utilizing the identical standardized test set (28,640 samples) released via Zenodo (DOI: 10.5281/zenodo.10145347).

This fixed, independent sample provides a controlled experimental framework that eliminates performance variance from stochastic data splitting and ensures direct comparability with the BNN baseline. By evaluating on this consistent dataset, we can isolate the architectural advantages of LSTM-MDNz in resolving complex color–redshift degeneracies and enhancing uncertainty quantification relative to current probabilistic methods.



\section{Methodology} \label{sec:method}
\subsection{The LSTM-MDNz Model}

The LSTM-MDNz model employed in this study is an end-to-end deep learning architecture designed to synergize the strengths of LSTM networks—a robust variant of recurrent neural networks (RNNs)—in sequential feature extraction with the advanced capability of MDNs to characterize multimodal and asymmetric probability distributions. This dual-component architecture is specifically engineered to resolve the inherent color–redshift degeneracies that frequently compromise photometric redshift estimation.

Previously, \citet{2026ApJS..282...46C} successfully applied the LSTM-MDNz framework to characterize the complex spectral energy distributions (SEDs) of quasars. By treating multi-band magnitudes as a physically motivated sequence, they achieved high-precision photometric redshift estimation and robust uncertainty quantification.

Building upon this foundation, the present study represents the first extension of this framework to large-scale galaxy samples. Unlike the relatively uniform power-law features of quasars, galaxy SEDs are governed by more intricate physical processes—including complex star formation histories (SFH), chemical enrichment, and varied dust attenuation. By generalizing the research focus from active galactic nuclei to this broader galaxy population, we aim to demonstrate the model’s efficacy in capturing subtle, wavelength-dependent features to resolve the color–redshift degeneracies prevalent in next-generation surveys like LSST and Euclid.

While a comprehensive mathematical derivation and implementation details are provided in \citet{2026ApJS..282...46C}, a concise overview of the model’s core logic and structural specifications as applied to galaxy photometry is provided below to ensure the self-containment of this work.

The LSTM-MDNz architecture is composed of two synergistic core modules. The front end is the LSTM module, centered around multiple layers of bidirectional long short-term memory (Bi-LSTM) networks. Unlike traditional neural networks that treat photometric bands as isolated features, this module interprets multi-band data as a wavelength-ordered sequence, treating the spectral energy distribution as a quasi-temporal flow. Through the specialized gating mechanisms of the LSTM—comprising the forget, input, and output gates—the model effectively captures the nonlinear evolution of the SED across wavelengths. 

In our concrete implementation, the front-end feature extraction consists of two stacked Bi-LSTM layers. The first Bi-LSTM layer comprises 128 hidden units per direction, generating a 256-dimensional concatenated output sequence, which is subsequently processed by the second Bi-LSTM layer with 64 hidden units per direction to yield a 128-dimensional hidden representation $\boldsymbol{h}$ \citep{2026ApJS..282...46C}. Each recurrent layer is coupled with a dropout rate of 0.25 and batch normalization to guarantee generalization stability. By processing this sequence of magnitude-error pairs $[\text{mag}_{\lambda}, \Delta\text{mag}_{\lambda}]$, the module extracts a high-dimensional hidden state vector $\boldsymbol{h}$, which serves as a deep representation of redshift-sensitive features. This approach inherently exploits inter-band feature gradients—such as the characteristic shifting of the 4000 Å break across different filter sets—providing significantly stronger physical representational power than simple feature concatenation \citep{2024MNRAS.535.1844L, 2026ApJS..282...46C}

Subsequently, the front-end features are passed to the MDN module. Unlike deterministic regression models that yield a single point estimate, the MDN reconstructs the posterior probability density function of the redshift by fitting a Gaussian mixture model:
\begin{equation}
P(z \mid X) = \sum_{k=1}^{K} \pi_k(X) \mathcal{N}\big(z \mid \mu_k(X), \sigma_k^2(X)\big),
\label{eq:mdn}
\end{equation}
where $X$ represents the input photometric sequence, and \ $\pi_k$, $\mu_k$, $\sigma_k$ denote the mixing coefficient, mean, and standard deviation of the $k$-th Gaussian component, respectively.

Here, for the $i$-th galaxy, although the raw input sequence $X_i$ is first transformed by the Bi-LSTM layers into a high-dimensional hidden representation $\boldsymbol{h}$ before reaching the mixture density network, the final GMM parameters ($\pi_k, \mu_k, \sigma_k$) remain deterministic composite functions of the initial observations. Expressing these parameters explicitly as functions of $X$ preserves the mathematically rigorous definition of the end-to-end conditional probability density under the negative log-likelihood optimization framework. 

In our concrete architectural layout, this continuous mapping is realized hierarchically: the vector $\boldsymbol{h}$ is passed through two sequential dense hidden layers (comprising 128 and 64 units, respectively) within the MDN module for non-linear abstraction, before reaching the final linear mapping layer. This configuration forms a three-layer fully connected topology within the MDN block dedicated to dynamically generating the parameters. Since this sequence-based framework operates efficiently on wavelength-ordered two-dimensional magnitude-error arrays, the entire architecture remains highly lightweight, containing a total of approximately $3.8 \times 10^5$ trainable parameters.

In this study, we set the number of Gaussian components to $K = 10$, a configuration that provides sufficient degrees of freedom to characterize complex PDF morphologies. The final fully connected layer directly outputs the $3K = 30$ distribution parameters (10 mixing coefficients, 10 means, and 10 standard deviations). This framework is capable of modeling not only standard unimodal distributions but also the non-Gaussian features—such as multimodality and pronounced asymmetry—that arise from inherent color–redshift degeneracies.

\subsection{Training Strategy}

The LSTM-MDNz model was developed using the TensorFlow framework and implemented via the Keras API. Model training was conducted on a computing server equipped with an NVIDIA L40S GPU. To optimize the predicted PDFs, we employed the negative log-likelihood (NLL) as the loss function:
\begin{equation}
\mathcal{L} = -\frac{1}{N} \sum_{i=1}^{N} \ln \left[ \sum_{k=1}^{K} \pi_k(X_i) \mathcal{N}\big(z_{\rm spec,i} \mid \mu_k(X_i), \sigma_k^2(X_i)\big) \right],
\label{loss_equ}
\end{equation}
where $N$ is the total number of galaxies in the training batch, and $z_{\rm spec,i}$ is the ground-truth spectroscopic redshift of the $i$-th sample. This loss function is specifically designed to maximize the likelihood of the true redshift within the predicted probability distribution.  We used the Adam optimizer \citep{2014arXiv1412.6980K} with an initial learning rate of $1 \times 10^{-3}$ and moment parameters $\beta_1 = 0.9$, $\beta_2 = 0.999$.

To ensure stable convergence in the complex NLL loss landscape inherent to MDNs, we adopted the ReduceLROnPlateau learning rate scheduler \citep{2012arXiv1206.5533B}. This strategy monitors the validation loss and dynamically adjusts the learning rate: if no improvement is observed for 10 consecutive epochs (patience period), the learning rate is reduced by a factor of 0.2—i.e., decayed to 20\% of its previous value—until it reaches a predefined lower bound of $1 \times 10^{-9}$. This adaptive approach balances global exploration in early training with fine-grained convergence in later stages, preventing the optimizer from oscillating around or overshooting the global minimum. By systematically reducing the step size upon encountering performance plateaus, the model better navigates the non-convex optimization landscape and achieves a more robust, generalized solution.

Furthermore, to prevent overfitting and ensure optimal generalization, we implemented an early stopping strategy. Training is automatically terminated if the validation loss fails to improve for 30 consecutive epochs, after which the model weights corresponding to the minimum validation loss are restored. A batch size of 128 was adopted to balance gradient stability and memory efficiency. Under the described hardware configuration, the training process typically converges within 30 minutes, with the total inference time for the entire dataset being approximately 1 minute. This high computational efficiency, coupled with rapid inference, underscores the scalability of our framework for processing the massive data streams expected from next-generation wide-field surveys.


\section{Results} \label{sec:results}

This section evaluates the performance of the LSTM-MDNz model for photometric redshift estimation on the GalaxiesML test set. Our assessment focuses on both the statistical accuracy of point estimates and the calibration and reliability of the predicted posterior PDFs. To facilitate reproducibility and support community research, the model architecture, trained weights, and evaluation scripts used in this study are publicly available in our GitHub repository\footnote{\url{https://github.com/zjluo-code/LSTM-MDNz-GalaxiesML}}.

\subsection{Evaluation Metrics}

The evaluation metrics in this study are categorized into two primary types: point-estimation statistics and PDF evaluation metrics. For each sample, we define the final photometric redshift estimate ($z_{\rm phot}$) as the expectation value of the predicted conditional distribution:
\begin{equation}
z_{\rm phot} = E[z \mid X] = \int z P(z \mid X) dz = \sum_{k=1}^{K} \pi_k(X) \mu_k(X).
\label{z_phot_equ}
\end{equation}

The point-estimation statistics are employed to assess the accuracy and dispersion of these $z_{\rm phot}$ values relative to their spectroscopic counterparts ($z_{\rm spec}$), focusing on the central tendency of the predictions. Complementarily, the PDF evaluation metrics provide a quantitative assessment of the calibration and statistical reliability of the posterior distributions, offering a rigorous measure of how well the model characterizes underlying redshift uncertainties.

Regarding point estimation, to ensure direct comparability with the results of \citet{2024ApJ...964..130J}, we adopt their core performance indicators, including the root mean square error (RMSE), outlier fraction ($O$), catastrophic outlier rate ($O_C$), scatter ($Scatter$), and median bias ($bias$). These metrics collectively characterize the accuracy and dispersion of the redshift point estimates, providing a standardized basis for benchmarking our architecture.

To further assess model robustness against non-Gaussian features and extreme outliers, we incorporate two additional metrics widely utilized in contemporary photometric redshift studies \citep{2018A&A...619A..14F,2020A&A...644A..31E,2021MNRAS.502.2770M,2024MNRAS.52712140L,2024MNRAS.531.3539L,2024MNRAS.535.1844L,2026ApJS..282...46C}: the mean absolute error (MAE) and the normalized median absolute deviation ($\sigma_{\rm NMAD}$). Although these metrics were not included in the initial evaluation suite by \citet{2024ApJ...964..130J}, their inclusion provides a more comprehensive characterization of performance, particularly given the sparse, high-redshift populations and heteroscedastic noise inherent in the GalaxiesML dataset.

Among these metrics, RMSE and MAE quantify the average deviation between the photometric and spectroscopic redshifts, normalized by the factor $(1+z_{\mathrm{spec}})$ to account for the intrinsic expansion of the error budget at higher redshifts. These normalized residuals are defined as:
\begin{equation}
\mathrm{RMSE} = \sqrt{\frac{1}{N} \sum_{i=1}^{N} \left(\frac{z_{\mathrm{phot},i} - z_{\mathrm{spec},i}}{1+z_{\mathrm{spec},i}}\right)^2},
\label{rmse_equ}
\end{equation}
\begin{equation}
\mathrm{MAE} = \frac{1}{N} \sum_{i=1}^{N} \frac{|z_{\mathrm{phot},i} - z_{\mathrm{spec},i}|}{1+z_{\mathrm{spec},i}},
\label{mae_equ}
\end{equation}
where $N$ denotes the total number of samples in the test set, and $i$ represents the index of each individual galaxy.

$O$ and $O_C$ quantify the proportions of samples exhibiting significant redshift estimation errors. Following the established criteria in the literature \citep[e.g.,][]{2018A&A...619A..14F, 2020A&A...644A..31E}, a photometric redshift estimate is classified as an outlier or a catastrophic outlier if it satisfies the following respective conditions:
\begin{equation}
\frac{|z_{\mathrm{phot},i} - z_{\mathrm{spec},i}|}{1 + z_{\mathrm{spec},i}} > 0.15,
\label{out_equ}
\end{equation}
\begin{equation}
|z_{\mathrm{phot},i} - z_{\mathrm{spec},i}| > 1.0.
\label{cat_out_equ}
\end{equation}
The metrics $O$ and $O_C$ represent the percentage of the total test sample meeting these criteria. While $O$ focuses on the relative error normalized by the redshift, $O_C$ identifies samples with absolute deviations exceeding a unit redshift, which are particularly detrimental to cosmological studies.

$Scatter$ and $bias$ quantify the statistical fluctuations and systematic offsets in the redshift estimates, respectively. Specifically, bias reflects the systematic tendency of the predictions to be higher or lower than the spectroscopic values, while $scatter$ characterizes the typical width of the error distribution. We define the normalized residual for the $i$-th sample as:
\begin{equation}
\Delta z'_i = \frac{z_{\mathrm{phot},i} - z_{\mathrm{spec},i}}{1 + z_{\mathrm{spec},i}}.
\label{resid_equ}
\end{equation}
These metrics are then calculated using robust estimators as follows:
\begin{equation}
bias = \mathrm{median} \left( \Delta z'_i \right),
\label{bias_equ}
\end{equation}
\begin{equation}
Scatter = \mathrm{median} \left( \left| \Delta z'_i - \mathrm{median}(\Delta z'_i) \right| \right).
\label{scatter_equ}
\end{equation}

Finally, we calculate $\sigma_{\rm NMAD}$, a robust dispersion metric that provides a complementary perspective to the $Scatter$ defined in Eq.~\ref{scatter_equ}. While both utilize median-based estimators to mitigate the impact of outliers, $\sigma_{\rm NMAD}$ applies the redshift-dependent scaling directly to the centered residuals. Following \citet{2008ApJ...686.1503B}, it is defined as:
\begin{equation}
\sigma_{\rm NMAD} = 1.4826 \times \mathrm{median} \left( \left| \frac{\Delta z_i - \mathrm{median}(\Delta z_i)}{1 + z_{\mathrm{spec},i}} \right| \right),
\label{sigma_nmad_equ}
\end{equation}
where $\Delta z_i = z_{\mathrm{phot},i} - z_{\mathrm{spec},i}$. Unlike the global $Scatter$, this formulation accounts for the expansion of the error volume at higher redshifts by scaling the centered physical residuals ($\Delta z_i - \mathrm{median}(\Delta z_i)$) for each individual source. The coefficient $1.4826$ ensures that $\sigma_{\rm NMAD}$ is equivalent to the standard deviation for a purely Gaussian distribution. This approach yields a more reliable measure of the model's intrinsic predictive precision, effectively insulating the evaluation from the non-Gaussian tails and catastrophic outliers that often contaminate deep photometric surveys.

Regarding PDF evaluation metrics, in addition to the PIT adopted by \citet{2024ApJ...964..130J}, this study introduces the CRPS to examine the consistency between predicted probability distributions and true observations \citep{2018A&A...609A.111D}. The CRPS is a strictly proper scoring rule that simultaneously assesses both the accuracy (calibration) and sharpness (concentration) of probabilistic forecasts. By comparing the cumulative distribution function (CDF) of the predicted PDF against the Heaviside step function centered at the spectroscopic observation, it yields a scalar score representing the overall predictive performance. For a predictive CDF $F$ and a true spectroscopic redshift $z_{\rm spec}$, the CRPS for a single sample is defined as:
\begin{equation}
\mathrm{CRPS}(F, x) = \int_{-\infty}^{+\infty} \left[ F(y) - \mathbf{1}_{\{y \geq x\}} \right]^2 dy,
\label{eq:crps_equ}
\end{equation}
where $y$ is the integration variable and $\mathbf{1}_{\{y \geq x\}}$ denotes the indicator (Heaviside) function. A lower CRPS value signifies a more reliable and precise probabilistic prediction.

The PIT is a diagnostic tool used to validate the calibration of probabilistic forecasts, assessing whether the predicted distributions are statistically consistent with the true underlying uncertainties \citep{dawid1984present, 2010MNRAS.406..881B, 2018PASJ...70S...9T, 2018A&A...609A.111D, 2020A&A...644A..31E, 2021MNRAS.502.2770M}. For each sample, the PIT value is defined as the CDF of the predicted PDF evaluated at the true spectroscopic redshift:
\begin{equation}
\mathrm{PIT} = \int_{-\infty}^{z_{\rm spec}} P(z \mid X) , dz,
\label{pit_equ}
\end{equation}
where $z_{\rm spec}$ is the spectroscopic redshift and $P(z \mid X)$ is the posterior PDF generated by the model. According to the criterion established by \citet{dawid1984present}, for a perfectly calibrated model, the aggregate PIT values across a representative sample should follow a standard uniform distribution $U(0,1)$.

Deviations from uniformity reveal specific systematic distortions in the ensemble of predicted PDFs. A U-shaped PIT distribution indicates that the predicted PDFs are under-dispersed (too narrow or over-confident), causing the true values to fall more frequently in the tails of the distributions than predicted. Conversely, a humped (inverted U-shaped) distribution suggests the PDFs are over-dispersed (too broad or under-confident). Furthermore, a skewed PIT distribution indicates a systematic bias (overestimation or underestimation) in the model's central predictions \citep{2016arXiv160808016P}.

\subsection{Quantitative Analysis of Point Estimation Accuracy} \label{subsec:point}

This subsection evaluates the point-estimation performance of the LSTM-MDNz model using the full GalaxiesML test set. To establish a rigorous performance benchmark, we retrieved the BNN-based photometric redshift point estimates provided by \citet{2024ApJ...964..130J} from the Zenodo platform (DOI: 10.5281/zenodo.10145347) to serve as a direct baseline for comparison.

Subsequently, we employed a unified evaluation pipeline to perform consistent metric calculations on the results of both models. This approach effectively eliminates potential systematic biases arising from discrepancies in algorithmic metric implementation, ensuring the impartiality and reproducibility of the comparative analysis between LSTM-MDNz and the BNN baseline. The detailed quantitative performance metrics are summarized in Table \ref{table:pointmetrics}.

\begin{table*}[t] 
\caption{Quantitative comparison of photometric redshift point estimation metrics across different models. Bold values indicate the best performance between the primary probabilistic models (BNN, LSTM-MDNz, and its ablation variant).}
\label{table:pointmetrics}
\centering
\setlength{\tabcolsep}{12pt}
\begin{tabular}{l c c c c c c c} %
\hline\hline
\noalign{\smallskip}
Model & $O$ (\%) & $O_C$ (\%) & $Scatter$ & $\sigma_{\rm NMAD}$ & $bias^a$ & RMSE & MAE \\
\noalign{\smallskip}
\hline
\noalign{\smallskip}
BNN& 7.9 & 2.3 & 0.026 & 0.026 & -0.0024 & 0.145 & 0.055 \\
LSTM-MDNz (Mags Only) & 7.7 & 2.2 & 0.025 & 0.025 & -0.0023 & 0.150 & 0.056 \\
LSTM-MDNz & \textbf{6.5} & \textbf{1.8} & \textbf{0.023} & \textbf{0.024} & \textbf{-0.0013} & \textbf{0.130} & \textbf{0.048} \\
\noalign{\smallskip}
\hline
\noalign{\smallskip}
Mizuki & 27.4 & 10.2 & 0.055 & -- & 0.011 & 0.307 & -- \\
DEmP & 25.0 & 9.2 & 0.040 & -- & 0.002 & 0.277 & -- \\
\noalign{\smallskip}
\hline
\noalign{\smallskip}
RF & 9.2 & 0.6 & 0.012 & -- & 0.001 & 0.088 & -- \\
XGBoost & 10.5 & 2.2 & 0.033 & -- & 0.002 & 0.149 & -- \\
SPIDERz & 9.0 & 5.1 & 0.044 & -- & 0.002 & 0.199 & -- \\
\noalign{\smallskip}
\hline
\end{tabular}
\tablefoot{$^a$ Bias definitions vary (Median Deviation for BNN/LSTM-MDNz vs. Mean Absolute Deviation for others).}
\end{table*}

To provide a comprehensive performance reference, Table~\ref{table:pointmetrics} incorporates several baseline results from \citet{2024ApJ...964..130J}, including the SVM-based SPIDERz \citep{2017A&A...600A.113J}, Random Forest \citep{2001MachL..45....5B}, and the gradient-boosting model XGBoost \citep{2016arXiv160302754C}. Critically, these machine learning baselines were evaluated on the same GalaxiesML test set as our model, ensuring a direct and consistent comparison. Additionally, benchmark results from the template-fitting code Mizuki \citep{2018PASJ...70S...9T} and the empirical method DEmP \citep{2014ApJ...792..102H} are included. It should be noted that the evaluation for these two methods was conducted on a cross-matched sub-sample of approximately 60,000 galaxies, constructed by \citet{2024ApJ...964..130J} using overlapping data from wider surveys \citep{2020arXiv200301511N}. Given that \citet{2024ApJ...964..130J} has already demonstrated that the BNN framework generally outperforms these traditional models across the full sample, our subsequent analysis focuses primarily on the detailed comparison between LSTM-MDNz and the BNN baseline.

As summarized in Table~\ref{table:pointmetrics}, the LSTM-MDNz model achieves systematic performance gains over the BNN baseline across all statistical indicators. In terms of point estimation accuracy, the RMSE decreases from 0.145 to 0.130 (a $\sim$10\% improvement), while the MAE drops from 0.055 to 0.048 (a $\sim$13\% improvement). Furthermore, the metrics characterizing the core dispersion of the predictive distribution, $\sigma_{\rm NMAD}$ and $Scatter$, exhibit measurable refinements, decreasing from 0.026 to 0.024 and 0.023, respectively.

To rigorously break down the relative contributions of our sequential architecture versus input data features, we perform a dedicated ablation test by training a magnitude-only variant, LSTM-MDNz (Mags Only), which completely excludes the cmodel magnitude uncertainties ($\Delta{\rm mag}_\lambda$). As presented in Table~\ref{table:pointmetrics}, even when deprived of these cmodel photometric uncertainty columns, this pure-architecture baseline still yields point estimates that remain highly consistent with or modestly superior to the reference BNN framework, maintaining a comparable robust scatter ($\sigma_{\rm NMAD} = 0.025$) and a stable outlier fraction ($O = 7.7\%$). This quantitative resilience establishes that the sequential network design itself carries an intrinsic, structural capacity to capture underlying SED signatures along the wavelength axis, rather than relying passively on data-engineering dividends. When these uncertainty features are reintroduced into the full model (the baseline LSTM-MDNz), they provide adaptive probabilistic constraints for lower-${\rm SNR}$ sources, triggering a beneficial synergistic effect that further drives down the global outlier fraction to $6.5\%$.

This consistent enhancement across both global and robust metrics underscores the efficacy of the LSTM architecture in extracting wavelength-ordered sequential features. Unlike standard MLPs, which typically treat input features as independent combinations, LSTMs are inherently designed to capture subtle evolutionary trends within SEDs \citep{2024MNRAS.535.1844L}. Consequently, this sequential modeling approach imposes more rigorous physical constraints on the redshift estimation process within the GalaxiesML dataset, laying the foundation for handling more challenging sources where traditional methods often fail.

Beyond the improvements in point-estimation accuracy, LSTM-MDNz demonstrates superior robustness in mitigating extreme estimation failures. The outlier fraction ($O$) decreases from 7.9\% to 6.5\% (an $\sim$18\% relative reduction), while the catastrophic outlier rate ($O_C$) is suppressed from 2.3\% to 1.8\% (a $\sim$22\% relative reduction). This significant suppression of extreme biases reflects the synergistic effect between sequential uncertainty-aware feature extraction and multimodal posterior modeling.

This enhanced robustness stems from an integrated architectural design: by incorporating per-band photometric uncertainties as explicit features, the LSTM layers can dynamically extract physical representations based on the signal-to-noise ratio (S/N). This provides the MDN backend with error-aware contextual information, enabling a more effective optimization of mixture coefficients and component widths. Such end-to-end synergy allows the model to identify underlying uncertainties in ambiguous or noisy data, subsequently suppressing spurious probability peaks or appropriately broadening the posterior distribution to encompass the true redshift. This coupled mechanism significantly mitigates the risk of catastrophic failures triggered by physical degeneracies or observational noise.

Furthermore, the median bias ($bias$) of LSTM-MDNz is notably low at $-0.0013$, nearly halving the $-0.0024$ offset observed in the BNN baseline. This near-zero bias indicates strong statistical neutrality across the investigated redshift range, a characteristic that is essential for minimizing systematic errors in high-precision cosmological parameter estimations.

   \begin{figure*}
        \centering
        \includegraphics[width=18cm]{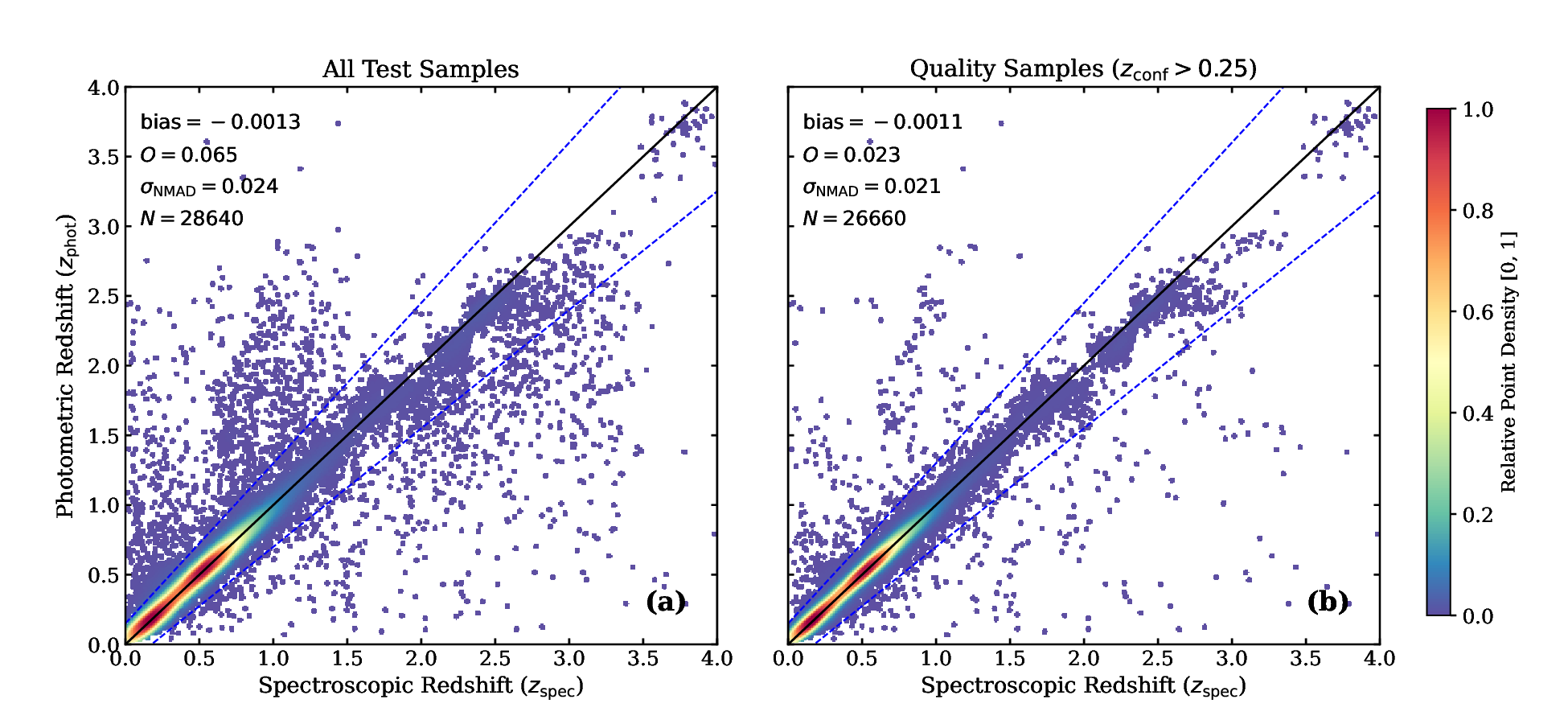}
        \caption{ Comparison between spectroscopic redshifts ($z_{\rm spec}$) and photometric redshifts ($z_{\rm phot}$) for the test set. Panel (a) shows the distribution for the full sample, while panel (b) displays the high-quality sample filtered by the PDF-based confidence metric ($z_{\rm conf} > 0.25$). The solid black line represents the $z_{\rm phot} = z_{\rm spec}$ diagonal, and the dashed blue lines on either side mark the outlier thresholds, defined as $|\Delta z| / (1+z_{\rm spec}) > 0.15$. The scatter color indicates the normalized density of the sample points, with the color gradient from blue to red representing increasing relative density.}
        \label{fig:zszp}
    \end{figure*}

Figure \ref{fig:zszp}(a) illustrates the 2D density distribution of photometric redshifts ($z_{\rm phot}$) against spectroscopic redshifts ($z_{\rm spec}$) for the test set. The vast majority of samples are tightly concentrated along the $z_{\rm phot} = z_{\rm spec}$ identity line, indicating high overall consistency. As visualized by the density contours and further supported by the global metrics in Table \ref{table:pointmetrics}, the model maintains a narrow dispersion ($\sigma_{\rm NMAD} = 0.024$) with negligible median bias. Notably, the bulk of the population remains well-contained within the $|\Delta z| \leq 0.15(1+z_{\rm spec})$ envelope (indicated by the dashed blue lines), with only a sparse fraction of catastrophic outliers ($O = 6.5\%$). This tight alignment across the entire redshift range ($0 < z < 4$) qualitatively confirms the capacity of the LSTM-MDNz architecture to capture the complex color–redshift mapping within the GalaxiesML dataset.

Regarding the global redshift correspondence, the model demonstrates high predictive fidelity in the low-to-intermediate regime ($z_{\rm spec} \lesssim 0.8$), where the high-density regions—represented by the warm red-to-yellow contours—tightly trace the diagonal. At higher redshifts ($z_{\rm spec} \gtrsim 0.8$), the distribution exhibits broader scatter. This trend is primarily attributed to the increased sparsity of training samples and the gradual migration of key spectral features, such as the $4000\,\text{\AA}$ break, beyond the HSC optical filters. Notably, while the core distribution maintains a clear linear trend, a subtle systematic underestimation becomes visible at $z_{\rm spec} \gtrsim 2.0$, manifested by a slight downward shift of the density peak. This behavior likely reflects the inherent challenge of resolving color–redshift degeneracies as galaxies become fainter and more reddened at high redshifts.

To mitigate the observed dispersion and systematic offsets in the high-redshift regime, we introduce a PDF-based confidence metric, $z_{\rm conf}$. While the formal definition and quantitative gains of $z_{\rm conf}$ are detailed in Section \ref{subsec:zconf}, its denoising and calibration effects are qualitatively illustrated in Figure \ref{fig:zszp}(b). By applying a representative threshold ($z_{\rm conf} > 0.25$), the model effectively suppresses outliers and rectifies distribution offsets at $z_{\rm spec} \gtrsim 0.8$, leading to a tighter alignment with the identity line. This comparison suggests that the LSTM-MDNz architecture possesses an internal filtering mechanism capable of isolating high-risk samples affected by observational limits. Such a capability is instrumental for constructing the high-purity galaxy catalogs required for precision cosmological analyses, such as weak lensing and large-scale structure studies.

\subsection{PDF Performance and Uncertainty Quantification} \label{subsec:pdf}

The scientific utility of photometric redshifts depends critically on the statistical calibration and reliability of the model-generated PDFs. Unlike traditional point-estimation methods, LSTM-MDNz yields a full posterior probability distribution for each galaxy, enabling a rigorous assessment of predictive uncertainty. In this subsection, we evaluate the probabilistic forecasting performance of the model using two complementary metrics: the CRPS for overall accuracy and sharpness, and the PIT for distributional calibration.

   \begin{figure*}
        \centering
        \includegraphics[width=18cm]{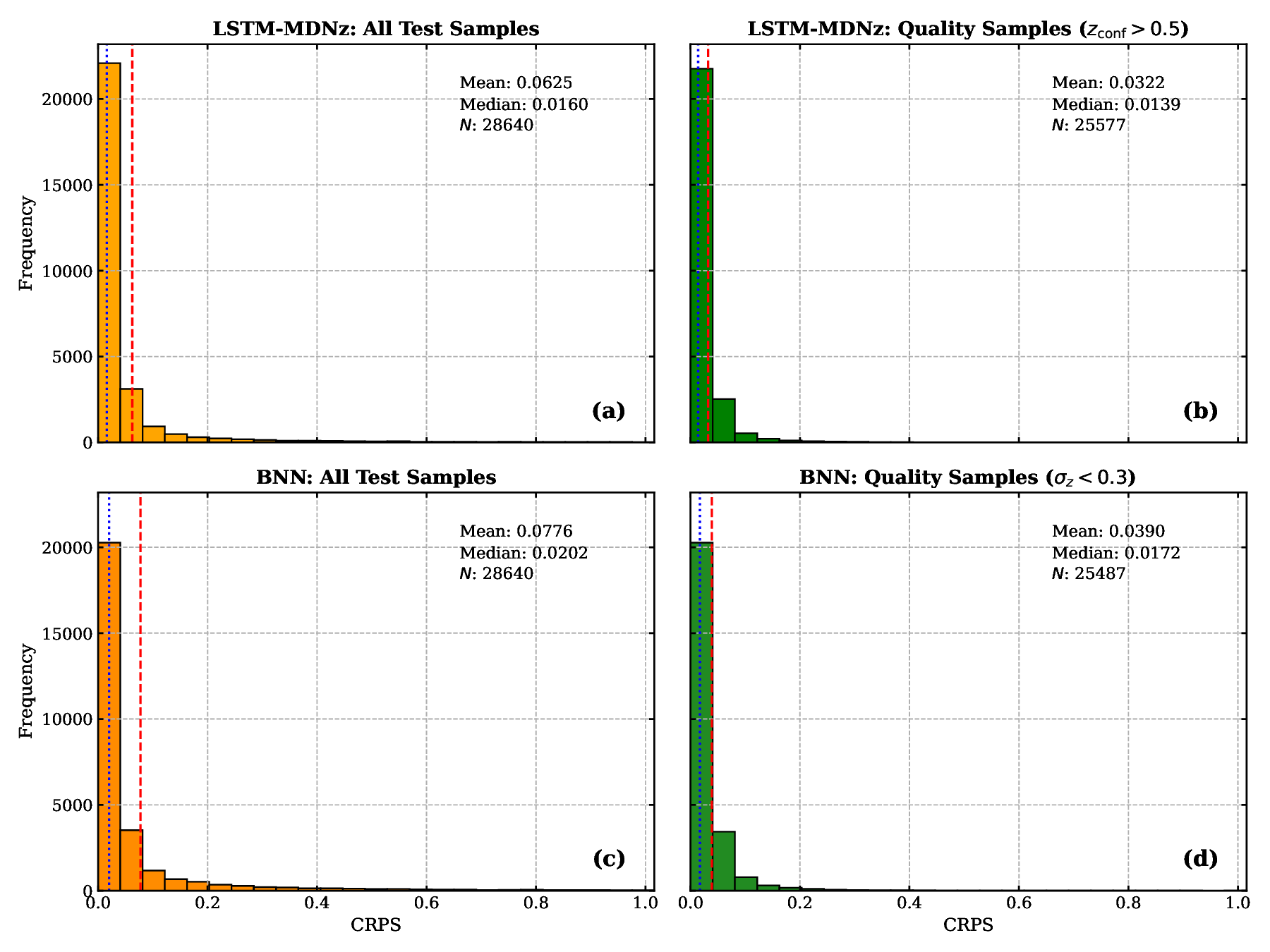}
        \caption{Comparison of CRPS distributions between the LSTM-MDNz model (top row) and the BNN model (bottom row) for the GalaxiesML test set. Panels (a) and (c) display the distributions for the full test samples of the respective models. Panel (b) shows the high-quality LSTM-MDNz sample filtered by the confidence metric ($z_{\rm conf} > 0.5$), while panel (d) shows the high-quality BNN sample selected by the predictive uncertainty threshold ($\sigma_{z} < 0.3$). In all panels, the red dashed line and the blue dotted line designate the mean and median CRPS values, respectively, and $N$ indicates the total number of galaxies within each sub-sample.}
        \label{fig:crps}
    \end{figure*}
    
The statistical distributions of CRPS values for the LSTM-MDNz model are shown in Figure \ref{fig:crps} along with the corresponding performance from the BNN baseline for comparison. As observed in the full test set distributions, the LSTM-MDNz model (Panel a) exhibits a median CRPS of 0.0160 with a mean of 0.0625, revealing a heavy tail caused by a small fraction of galaxies with large estimation errors. In comparison, the BNN baseline (Panel c) yields a higher median CRPS of 0.0202 and a mean of 0.0776. These metrics demonstrate that LSTM-MDNz achieves a systematic improvement over the BNN baseline, reducing the median and mean CRPS by approximately 21\% and 19\%, respectively, thus reflecting a better overall calibration and sharpness in its predicted PDFs. Given that CRPS shares the same dimensions as redshift ($z$), these median values indicate that for the vast majority of the galaxy population, both architectures are capable of generating posterior PDFs that maintain reliable probabilistic consistency with the spectroscopic ground truths.

However, for both models, the mean CRPS values are higher than their respective medians, reflecting right-skewed distributions with extended tails. This statistical characteristic indicates that the global average performance is largely influenced by a minority subset of challenging samples—typically those affected by severe color–redshift degeneracies or subtle spectral feature shifts.

To address these higher-error tails, we implement model-specific selection criteria designed to identify and isolate less constrained predictions. As illustrated in the right column of Figure \ref{fig:crps}, applying a PDF-based confidence threshold ($z_{\rm conf} > 0.50$) for the LSTM-MDNz model effectively suppresses the high-error distribution, lowering its mean CRPS from 0.0625 to 0.0322 (Panel b). Similarly, utilizing the predictive variance filtering ($\sigma_z < 0.3$) for the BNN baseline yields a comparable structural shift, reducing its mean CRPS from 0.0776 to 0.0470 (Panel d). A detailed quantitative evaluation of these filtering thresholds, along with the formal mathematical definition of the confidence metrics, is deferred to Section \ref{subsec:zconf} to maintain a progressive discussion.

The comparison across this two-by-two grid indicates that while the LSTM-MDNz model exhibits a higher overall probabilistic precision compared to the BNN baseline across both the full and filtered samples, both filtering mechanisms function as reliable indicators of predictive quality. This provides an effective approach for future deep surveys to construct higher-purity galaxy sub-samples by filtering out samples with degenerate or poorly constrained PDFs.
 
   \begin{figure*}
        \centering
        \includegraphics[width=18cm]{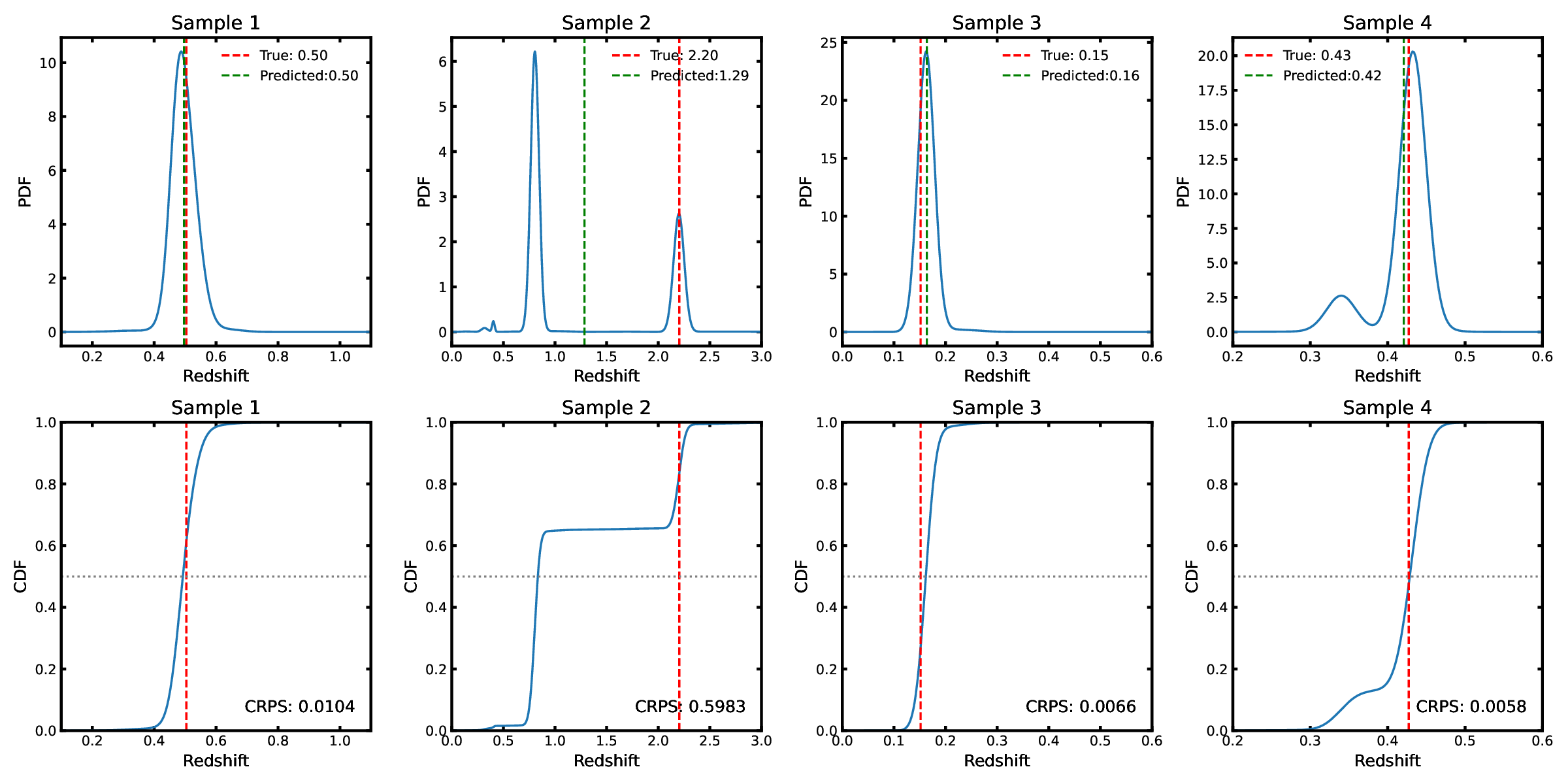}
        \caption{Examples of predicted photometric redshift PDFs and CDFs for four randomly selected typical samples. The top panel of each sub-figure displays the model-generated PDF, while the bottom panel shows the corresponding CDF. The red vertical dashed lines indicate the spectroscopic redshift ($z_{\rm spec}$), and the green vertical dashed lines represent the point estimates derived from the predicted PDF (i.e., $z_{\rm phot}$).}
        \label{fig:cdf}%
    \end{figure*}
    
Figure \ref{fig:cdf} illustrates the predicted PDFs (top panels) and their corresponding CDFs (bottom panels) for four representative test samples evaluated by the LSTM-MDNz model. The red dashed lines indicate the spectroscopic redshifts ($z_{\rm spec}$), while the green dashed lines represent the point estimates derived from the model. For most of these samples, the model-generated PDFs exhibit remarkably sharp, unimodal distributions, with the spectroscopic redshift aligning closely with the predictive peak. The corresponding CDFs show a near-step-function rise at the true redshift, yielding very low CRPS values (e.g., 0.0066 for Sample 3), which visually underscores the model’s high predictive sharpness and its capacity to yield accurate point estimates consistent with the ground truth.

Sample 2 in Figure \ref{fig:cdf} highlights a classic failure mode in deterministic point estimation driven by color–redshift degeneracy. In this instance, the posterior PDF generated by the LSTM-MDNz model manifests as a distinct multimodal distribution, with significant probability mass allocated to two physically plausible solutions at $z \approx 0.8$ and $z \approx 2.2$. The point estimate (green dashed line), calculated as the expected value of the full posterior, is positioned within the low-density region between the two discrete peaks. Due to this averaging effect, the predicted $z_{\rm phot}$ corresponds to a redshift where the actual probability density is nearly zero—a solution that is both statistically improbable and physically unmotivated.

This misalignment results in a relatively high CRPS (0.5983), reflecting the heightened predictive uncertainty inherent in such degeneracies. Notably, the spectroscopic redshift (red dashed line) is accurately captured within the high-probability region of the secondary peak. This multimodality further manifests as a distinct 'staircase' pattern in the corresponding CDF, where each horizontal plateau represents a region of negligible probability between competing hypotheses. This demonstrates that, unlike unimodal baseline models that would collapse these conflicting solutions into a single erroneous Gaussian, LSTM-MDNz explicitly preserves the topological complexity of the parameter space. By maintaining significant probability density at the true redshift, the model provides a self-consistent mechanism to flag predictive risks, thereby mitigating the propagation of systematic biases into subsequent cosmological analyses.

   \begin{figure*}
        \centering
        \includegraphics[width=18cm]{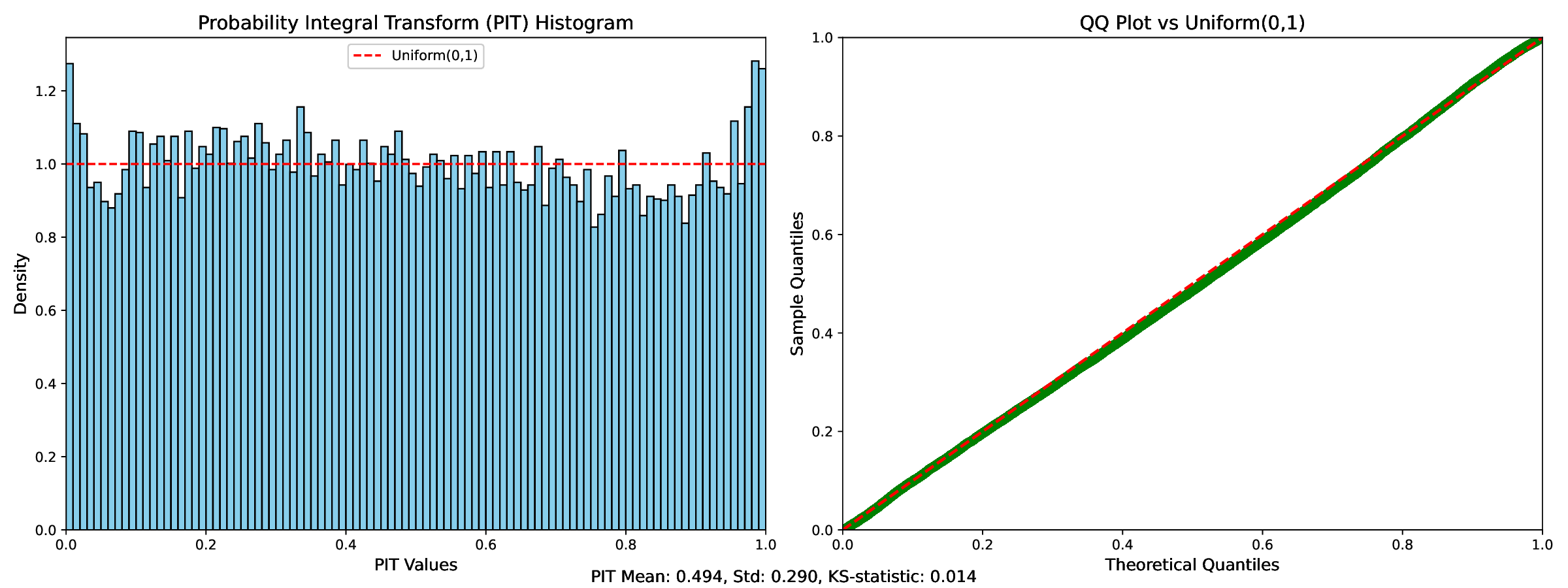}
        \caption{Assessment of the calibration quality of the model-predicted PDFs. Left panel: PIT distribution histogram for the full test set. The blue histogram represents the observed PIT density, and the red horizontal dashed line indicates the reference for an ideal uniform distribution, $U(0,1)$. If the predicted PDFs are well-calibrated, the PIT values should follow a standard uniform distribution. Right panel: Corresponding Q-Q plot of the PIT values. The green dots represent the correspondence between the observed quantiles and the theoretical uniform distribution quantiles, and the red diagonal dashed line is the theoretical reference ($y=x$). If the predicted PDFs are perfectly calibrated, all points should lie closely along this line.}
        \label{fig:pit}
    \end{figure*}

The PIT distribution of the test set obtained by the LSTM-MDNz model is presented in the left panel of Figure \ref{fig:pit}. The uniformity of this distribution serves as a diagnostic for the statistical calibration between the predicted uncertainties and the true observational errors. For a perfectly calibrated PDF, the PIT values follow a standard uniform distribution, $U[0,1]$ (indicated by the red horizontal dashed line). Our results show that the observed PIT distribution (blue histogram) closely aligns with this theoretical ideal, yielding a mean of 0.494 and a standard deviation of 0.290—approximating the expected values of 0.5 and 0.2886, respectively. Furthermore, the Kolmogorov–Smirnov (KS) statistic of 0.014 indicates a high degree of consistency with a uniform distribution.

In contrast, the baseline BNN model \citep{2024ApJ...964..130J}, which assumes a single Gaussian posterior, exhibits a slight inverted U-shape (humped) in its PIT distribution. This central concentration suggests that the BNN-generated PDFs may be over-dispersed (under-confident), leading to an overestimation of the underlying uncertainties. Conversely, the LSTM-MDNz distribution remains notably flat, exhibiting neither the U-shaped features indicative of under-dispersion nor the inverted U-shape characteristic of over-dispersion. This near-uniformity suggests that the overall scales of our predicted PDFs—whether unimodal or multimodal—accurately characterize the aleatoric uncertainty of the redshift estimates. Such rigorous calibration is important for downstream cosmological analyses that rely on well-characterized individual uncertainties. For instance, in frameworks where the global ensemble redshift distribution, $n(z)$, is estimated via PDF stacking, having accurate error scales helps maintain the statistical consistency needed to mitigate systematic biases.

The right panel of Figure \ref{fig:pit} presents the quantile–quantile (Q–Q) plot of the PIT values against a theoretical uniform distribution $U[0,1]$. For a well-calibrated posterior PDF, the PIT distribution should be uniform, causing the sample quantiles to align closely with the red diagonal identity line. As illustrated, the observed quantiles (green dots) exhibit strong agreement with the theoretical reference, with no significant systematic deviations even at the distributional extremes. This linear alignment provides independent evidence that the predictive uncertainties yielded by LSTM-MDNz are statistically robust and consistently calibrated across the entire redshift range. 

\subsection{High-Quality Sample Selection Based on $z_{\rm conf}$} \label{subsec:zconf}

From the perspective of the core cosmological goals of the LSST, the precision of photometric redshifts is fundamental to achieving accurate measurements of dark energy and cosmic shear. Within its most scientifically critical redshift range ($0.3 < z < 3.0$), the LSST survey demands not only a massive sample size (approximately 4 billion galaxies) but also robust algorithms capable of systematically identifying outliers and quantifying their impact on cosmological parameter estimation.

These stringent requirements pose a significant challenge for existing uncertainty quantification frameworks. Recent strategies for outlier mitigation have evolved along two primary paths: PDF morphology analysis and machine learning classification. Works such as \citet{2020PASP..132b4501J} and \citet{2021PASP..133d4504W} focus on the geometric properties of the posterior distribution, utilizing features of multimodality—such as the separation and relative probability mass of secondary peaks—to flag catastrophic errors. In contrast, \citet{2024ApJ...964..130J} utilize the predictive uncertainty ($\sigma_z$) from BNNs as a filtering benchmark; however, their framework typically assumes a unimodal Gaussian functional form for the redshift PDF, where $\sigma_z$ represents only a localized dispersion. This parametric limitation is inherently incapable of capturing topological risks, such as multimodal degeneracies or significant skewness, that arise from physical overlaps in color–magnitude space.

Parallel to these PDF-based methods, an alternative approach treats outlier identification as a binary classification task. \citet{2022ApJ...928....6S} demonstrated that neural networks can learn to recognize outlier-prone photometric patterns independently of PDF information, while \citet{2026ApJ...998..258Y} achieved a better trade-off between flagging efficiency and sample completeness by optimizing redshift-weighted loss functions. Nevertheless, such classification methods often face a generalization bottleneck: they typically require retraining for specific survey depths or redshift ranges. Moreover, because they output discrete labels rather than a continuous physical confidence score, they lack flexibility when integrated into complex downstream cosmological analysis pipelines.

To address these challenges, this study employs a confidence metric, $z_{\rm conf}$, derived from the morphological features of the predicted PDFs. Leveraging the capacity of LSTM-MDNz to model complex, multi-component posterior distributions, we capture key information in the non-Gaussian PDF to identify potential failure modes. Previous research has demonstrated that features such as multimodality and skewness are essential for distinguishing physical degeneracies \citep{2016MNRAS.457.4005W, 2018A&A...609A.111D, 2020A&C....3000362D, 2020PASP..132b4501J}. By synthesizing these PDF characteristics, $z_{\rm conf}$ maps complex distributional topologies into a single, robust selection criterion, effectively mitigating the contamination of cosmological signals by degenerate samples while maintaining high catalog purity.

Following established protocols for PDF-based catalog purification \citep{2013MNRAS.432.1483C}, we define $z_{\rm conf}$ as the integral of the PDF over the interval $[z_{\rm mean} - \delta, z_{\rm mean} + \delta]$. To account for the redshift-dependent growth of estimation uncertainty, the error window half-width is defined as $\delta = \alpha(1 + z_{\rm mean})$, where $\alpha$ approximates the intrinsic scatter of the algorithm. In this analysis, we adopt a baseline value of $\alpha = 0.05$. Robustness tests conducted across a range of $\alpha \in [0.03, 0.15]$ indicate that while the absolute $z_{\rm conf}$ values scale with $\alpha$, the resulting catalog selection and statistical conclusions remain insensitive to this choice. Consistent selection results can be maintained by adjusting the confidence thresholds relative to the specific $\alpha$ employed.

Physically, $z_{\rm conf}$ quantifies the concentration and unimodality of the posterior PDF. Galaxies with sharp, highly localized distributions yield higher $z_{\rm conf}$ values, indicating that the predictive probability mass is effectively concentrated within the expected redshift interval. Conversely, galaxies exhibiting multimodal or diffuse distributions—typical signatures of color–redshift degeneracies—result in lower $z_{\rm conf}$ values as the probability mass disperses across multiple peaks or broad wings. Unlike filtering strategies based solely on primary-to-secondary peak ratios or fixed-width intervals, $z_{\rm conf}$ accounts for the global topology of the posterior by integrating the local probability mass. This provides a more holistic representation of predictive uncertainty that is sensitive to both the spread and the secondary features of the distribution. Ultimately, this framework provides a physically motivated mechanism for outlier mitigation, enabling a systematic trade-off between catalog purity and sample completeness.

\begin{table*}[t]
\caption{Performance evaluation under different selection criteria for LSTM-MDNz and the BNN baseline.}
\label{table:zconf_metrics}
\centering
\begin{tabular}{l c c c c c c c c c c}
\hline\hline
\noalign{\smallskip}
Model & Selection criteria & $f_{\rm ret}$ (\%) & $O$ (\%) & $O_C$ (\%) & $Scatter$ & $\sigma_{\rm NMAD}$ & $bias$ & RMSE & MAE & CRPS \\
\noalign{\smallskip}
\hline
\noalign{\smallskip}
LSTM-MDNz & --- & 100 & 6.5 & 1.77 & 0.023 & 0.024 & $-0.0013$ & 0.130 & 0.048 & 0.063 \\
& $z_{\rm conf} > 0.05$ & 95.9 & 3.4 & 0.98 & 0.022 & 0.022 & $-0.0011$ & 0.102 & 0.036 & 0.047 \\
& $z_{\rm conf} > 0.15$ & 94.3 & 2.6 & 0.70 & 0.022 & 0.021 & $-0.0011$ & 0.087 & 0.032 & 0.041 \\
& $z_{\rm conf} > 0.25$ & 93.1 & 2.3 & 0.59 & 0.021 & 0.021 & $-0.0011$ & 0.080 & 0.030 & 0.038 \\
& $z_{\rm conf} > 0.30$ & 92.4 & 2.1 & 0.55 & 0.021 & 0.021 & $-0.0011$ & 0.078 & 0.029 & 0.037 \\
& $z_{\rm conf} > 0.50$ & 89.3 & 1.6 & 0.39 & 0.020 & 0.020 & $-0.0011$ & 0.067 & 0.026 & 0.032 \\
& $z_{\rm conf} > 0.60$ & 86.9 & 1.3 & 0.33 & 0.020 & 0.019 & $-0.0011$ & 0.063 & 0.024 & 0.030 \\
\noalign{\smallskip}
\hline
\noalign{\smallskip}
BNN & --- & 100 & 7.9 & 2.30 & 0.026 & 0.026 & $-0.0024$ & 0.145 & 0.055 & 0.078 \\
& $\sigma_z < 0.5$ & 92.3 & 3.4 & 0.71 & 0.023 & 0.023 & $-0.0024$ & 0.085 & 0.035 & 0.047 \\
& $\sigma_z < 0.3$ & 89.0 & 2.4 & 0.45 & 0.022 & 0.022 & $-0.0024$ & 0.074 & 0.030 & 0.039 \\
\noalign{\smallskip}
\hline
\end{tabular}
\tablefoot{$f_{\rm ret}$ denotes the fraction of samples retained after applying the selection criteria. Reported CRPS values represent the mean CRPS for the corresponding subsamples.}
\end{table*}

Table~\ref{table:zconf_metrics} presents the performance metrics of LSTM-MDNz and the BNN baseline model under different selection criteria, along with the corresponding sample retention fractions ($f_{\rm ret}$). The results indicate that as the $z_{\rm conf}$ threshold increases, the purity of the output catalog exhibits a systematic monotonic improvement.

Specifically for the LSTM-MDNz model, a preliminary cut of $z_{\rm conf} > 0.05$ alone is sufficient to reduce the outlier rate ($O$) significantly from 6.5\% to 3.4\%, while maintaining a sample completeness as high as 95.9\%. When the selection is tightened to $z_{\rm conf} > 0.25$, the outlier fraction is further suppressed to 2.3\%, with the sample retention fraction still standing at 93.1\%. Under the most stringent threshold ($z_{\rm conf} > 0.60$), the outlier rate ($O$) drops to a very low 1.3\%, while sample completeness remains at 86.9\%. Moreover, precision metrics such as $O_C$, RMSE, and MAE also exhibit consistent improvement. This level of performance is sufficient to readily meet the stringent outlier control requirements of large-scale projects like LSST, while retaining the vast majority of the survey sample.

It is worth emphasizing that $z_{\rm conf}$ does not improve precision by correcting the point estimates of individual redshifts; rather, it serves as a selective filter that identifies and discards samples with lower predictive reliability. The data analysis in the table shows that the changes in core dispersion metrics (e.g., $\sigma_{\rm NMAD}$ decreasing from 0.024 to 0.019, $Scatter$ from 0.023 to 0.020) are relatively modest, indicating that the core sample population retained after $z_{\rm conf}$ selection maintains a high degree of statistical consistency. However, the average CRPS metric drops significantly by approximately 52\% (from 0.063 to 0.030), providing strong evidence that the mechanism can accurately identify and remove samples with low-quality PDFs. These excluded galaxies typically exhibit broadened or multi-modal distribution shapes, characteristic of typical color--redshift degeneracies. By filtering out such high-risk cases, this mechanism substantially enhances the global statistical purity of the catalog without requiring recalibration of the underlying model.

To evaluate the relative performance of LSTM-MDNz, we compare its selection behavior with the BNN baseline strategy established by \citet{2024ApJ...964..130J}, which relies on the predictive uncertainty $\sigma_z$. The results indicate that at similar sample retention fractions, the LSTM-MDNz framework exhibits a more pronounced purification efficiency and improved probabilistic calibration. 

For instance, at a moderate filtering level, using a $z_{\rm conf} > 0.30$ threshold allows our model to retain 92.4\% of the sample while maintaining an outlier fraction of 2.1\% and a mean CRPS of 0.037. In comparison, the BNN baseline at a comparable retention level ($\sigma_z < 0.5$; retention 92.3\%) yields an outlier rate of $O = 3.4\%$ and a mean CRPS of 0.047. Under more stringent purity requirements, a cut of $z_{\rm conf} > 0.50$ yields a retention fraction of 89.3\% with an outlier rate of 1.6\% and a further reduced CRPS of 0.032. In contrast, when the BNN is subjected to a tighter constraint ($\sigma_z < 0.3$), it retains a similar proportion of the sample (89.0\%) but results in an outlier rate of 2.4\% and a CRPS score of 0.039.

These systematic comparisons, both in terms of outlier reduction and global CRPS metrics, indicate that the $z_{\rm conf}$ mechanism offers a highly viable approach for next-generation cosmological surveys. By explicitly accounting for the full profile of the probability density function rather than relying solely on the distribution width ($\sigma_z$), this approach provides a well-balanced trade-off between discriminative power and sample utility.

After applying the $z_{\rm conf}$ selection, the improvement in model performance is further corroborated by the visual evolution of the $z_{\rm spec}$ vs. $z_{\rm phot}$ distribution. As shown in Figure \ref{fig:zszp}(b), the point estimation accuracy for the high-quality subset increases significantly; compared to the full sample distribution (Figure \ref{fig:zszp}a), the diffuse features in the $z \gtrsim 0.8$ region, triggered by color–redshift degeneracies, are substantially reduced. In particular, catastrophic outliers deviating significantly from the diagonal are effectively suppressed, making the overall distribution highly converged toward the identity line. This confirms that $z_{\rm conf}$ can not only sensitively identify mispredictions caused by multimodal posteriors, but also significantly enhance the statistical reliability and physical consistency of the remaining estimates by eliminating physically degenerate samples.

\begin{figure}[ht!]
   \centering
   \includegraphics[width=9cm]{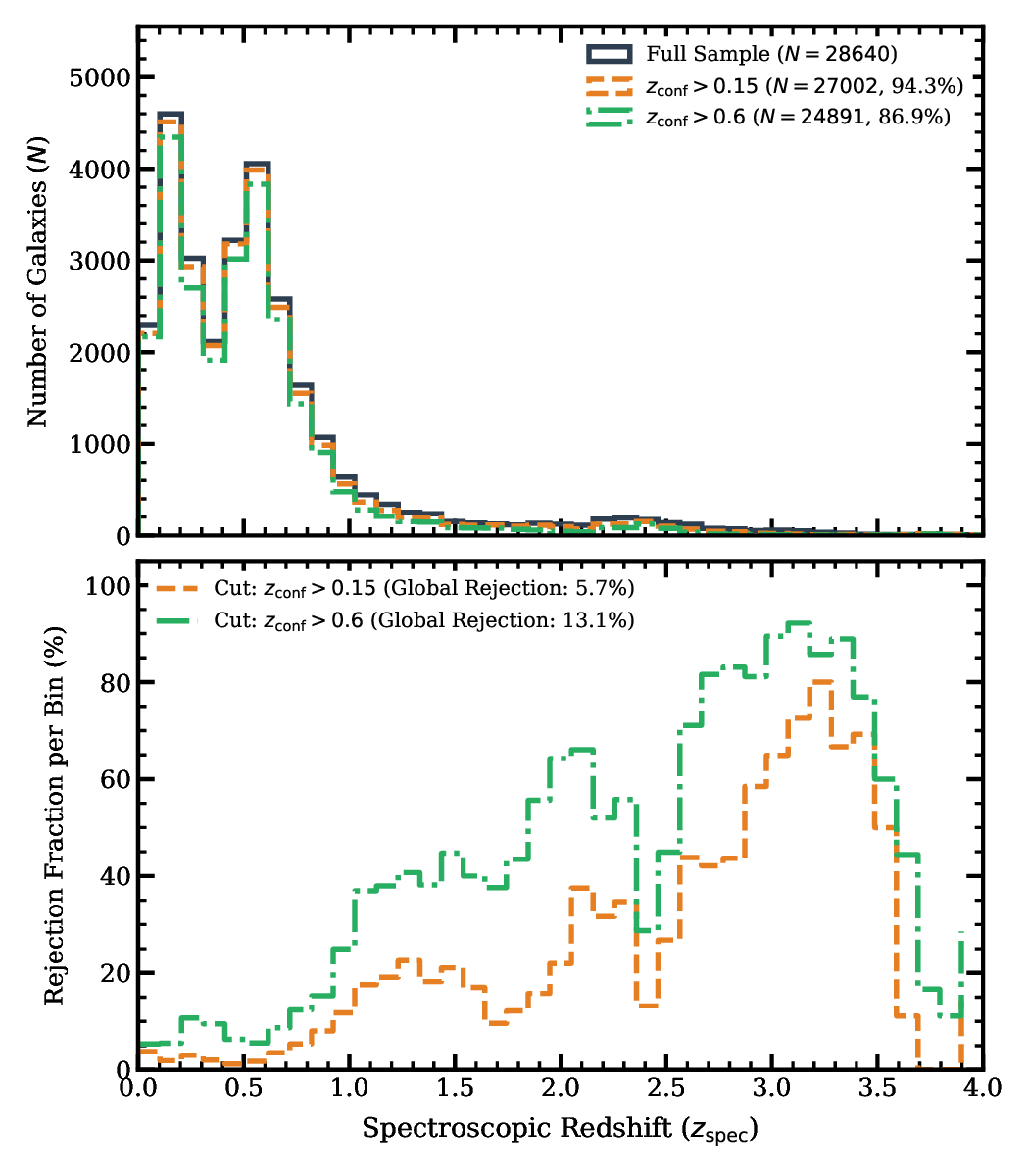}
   \caption{Top Panel: Evolution of the true (spectroscopic) redshift distribution $n(z_{\rm spec})$ under various $z_{\rm conf}$ selection thresholds. Bottom Panel: Fraction of rejected galaxies within each spectroscopic redshift bin for two different confidence thresholds ($z_{\rm conf} > 0.15$ and $z_{\rm conf} > 0.6$).}
   \label{fig:rej_frac}
   \end{figure}

To systematically understand the physical and statistical nature of galaxy filtering by the confidence metric $z_{\rm conf}$, we analyze the resulting spectroscopic redshift distribution $n(z_{\rm spec})$ of the subsamples retained under different $z_{\rm conf}$ thresholds, along with the corresponding rejection fractions across discrete redshift bins, as illustrated in Figure \ref{fig:rej_frac}. For low-redshift galaxies ($z_{\rm spec} \lesssim 0.8$), the $z_{\rm conf}$ selection cut has a minimal impact; for example, when applying a threshold of $z_{\rm conf} > 0.15$, less than 5\% of sources are filtered out (see Figure \ref{fig:rej_frac}, Bottom Panel). Rejections are primarily concentrated around $z_{\rm spec} \sim 1.2$, $z_{\rm spec} \sim 2.1$, and within the interval $2.6 \lesssim z_{\rm spec} \lesssim 3.5$. These correspond to critical cosmic epochs where spectral break features (Balmer or Lyman) either migrate directly across broadband filter boundaries or give rise to the classic long-range Balmer/Lyman break degeneracy. These are the typical redshift regimes that produce severe color-redshift degeneracies and lead to inherently multimodal posterior PDFs. 

Beyond these discrete physical degeneracy zones, the selection behavior of $z_{\rm conf}$ is found to be continuously modulated by the photometric data quality. Lower signal-to-noise ratios (${\rm SNR}$) naturally introduce larger flux measurement uncertainties, which subsequently increase the predictive uncertainty in the network outputs. The presence of such observational noise typically broadens the predicted redshift PDFs, leading to a systematic reduction in the corresponding $z_{\rm conf}$ values. This statistical behavior is reflected in the $i$-band cmodel measurements: the full test sample exhibits a median statistical SNR (${\rm SNR_{stat}}$) of 738, while the subset retained under a $z_{\rm conf} > 0.6$ threshold shows an increased median ${\rm SNR_{stat}}$ of 806. This monotonic trend illustrates that the $z_{\rm conf}$ metric acts as a viable probabilistic indicator linked to both geometric color-space degeneracies and data statistical quality, thereby helping to select a more reliable sample subset for downstream cosmological analysis.

Concurrently, the statistical distribution of CRPS in Figure \ref{fig:crps}(b) exhibits a notable increase in concentration. The heavy-tail characteristic observed in the full sample (Figure \ref{fig:crps}b) is markedly suppressed after selection, as evidenced by the mean (red dashed line) shifting closer to the median (blue dashed line). This morphological shift in the CRPS distribution indicates that the influence of extreme outliers has been significantly mitigated, resulting in a more refined subset where the PDFs are generally more compact and well-calibrated. Consequently, this improvement enhances the overall predictive reliability for subsequent cosmological applications.

\subsection{Performance Evaluation Across Redshift Bins}

To evaluate the model’s generalization and robustness across varying cosmic depths, we partitioned the test set into discrete redshift bins. Because the fidelity of the redshift distribution $N(z)$ within each slice is critical for minimizing systematic biases in cosmological inference, we assessed whether our photometric redshifts maintain consistent statistical accuracy and predictive precision across the core observational range ($0 < z_{\rm spec} < 2.5$). Bin-wise validation prevents localized performance issues—such as systematic offsets in specific redshift regimes—from being hidden by global statistical averages.

   \begin{figure*}
        \centering
        \includegraphics[width=18cm]{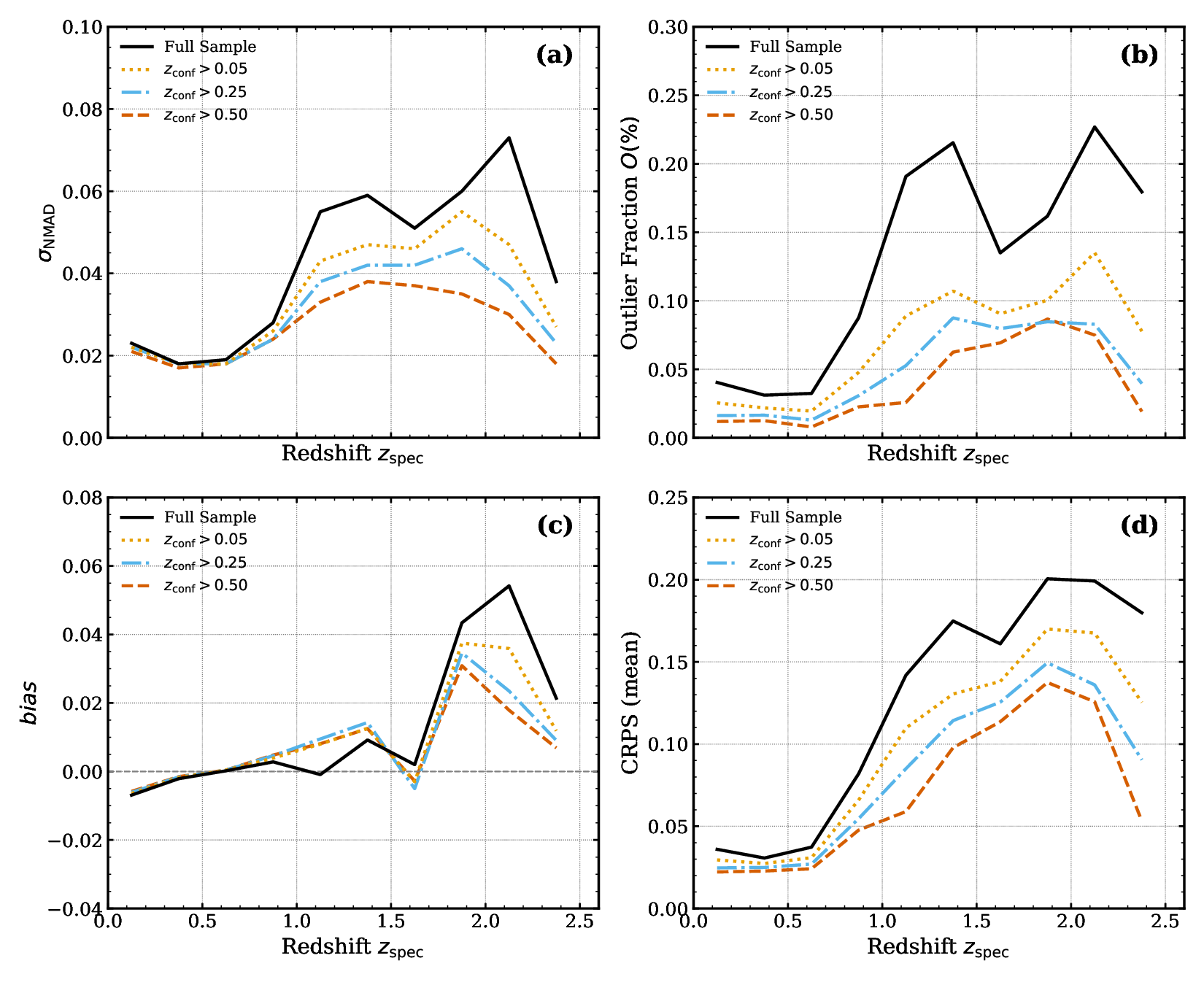}
        \caption{Bin-wise evolution of photometric redshift performance metrics as a function of spectroscopic redshift ($z_{\rm spec}$). The solid black line represents the full test set, while the colored dashed lines (see legend) correspond to subsets filtered by different $z_{\rm conf}$ thresholds. The four panels are: (a) $\sigma_{\rm NMAD}$; (b) outlier fraction $O (\%)$; (c) $bias$; and (d) mean CRPS.}
        \label{fig:zbin}%
    \end{figure*}
    
This interval-based evaluation is especially critical at high redshifts ($z \gtrsim 0.8$). As illustrated in Figure \ref{fig:z_dis}, the observed galaxy number density declines sharply with increasing redshift, limited by survey depth and the intrinsic scarcity of spectroscopic labels. In a volume‑averaged assessment, the high precision driven by the abundant low‑redshift population would otherwise mask significant degradation in the high‑$z$ regime, where signal‑to‑noise ratios (S/N) are lower and color–redshift degeneracies are most severe. For deep learning models, the ability to maintain stable predictive accuracy and generate physically motivated PDFs in these data‑sparse bins serves as a definitive benchmark of robustness against sample bias and readiness for high‑stakes scientific applications.

Figure~\ref{fig:zbin} illustrates the performance evolution of key photometric redshift metrics evaluated across discrete bins of spectroscopic redshifts, $z_{\rm spec}$, under varying $z_{\rm conf}$ selection thresholds. For the unfiltered full sample (solid black line), the normalized median absolute deviation ($\sigma_{\rm NMAD}$), the outlier fraction ($O$), and the mean CRPS all exhibit pronounced performance degradation beyond $z \approx 0.8$ (Figures~\ref{fig:zbin}a, b, and d). This trend reflects the inherent challenges of high-redshift estimation: sample sparsity, exacerbated color–redshift degeneracies, and the migration of key spectral features (e.g., the $4000\text{\AA}$ break) out of the observed filter passbands, which introduces ambiguities in color-space mapping. Notably, the metrics for the full sample show distinct peaks near $z \approx 1.3$ and $z \approx 2.1$, likely corresponding to sensitive regimes where major spectral features transition between filter bands.

The application of the $z_{\rm conf}$ selection mechanism leads to systematic and monotonic performance improvements across the entire redshift range. In the challenging $z \gtrsim 0.8$ regime, both $\sigma_{\rm NMAD}$ and $O$ undergo a substantial downward shift as the $z_{\rm conf}$ threshold increases. This demonstrates that $z_{\rm conf}$ effectively identifies and purges "high-risk" predictions—those fundamentally compromised by physical degeneracies—thereby significantly stabilizing the model's inference performance even in data-sparse regions.

Furthermore, the decline in mean CRPS (Figure~\ref{fig:zbin}d) with increasing $z_{\rm conf}$ closely mirrors the refinement of $\sigma_{\rm NMAD}$ and outlier rates. This cross-metric consistency suggests that the gain in point-estimation precision is primarily driven by the systematic truncation of heavy-tailed and multimodal structures within the posterior PDFs, consistent with the morphological evolution shown in Figure~\ref{fig:crps}(b). By isolating these compact and well-behaved posteriors, the $z_{\rm conf}$ framework significantly enhances the statistical fidelity of the sample, ensuring that the point estimates are supported by a rigorous probabilistic interpretation.

Regarding bias control, Figure~\ref{fig:zbin}c reveals that all subsets maintain near-zero bias for $z \lesssim 0.8$. At higher redshifts (particularly $z \gtrsim 1.5$), while the bias of the full sample fluctuates considerably (peaking near $z \approx 2.1$), the stringent selection subset (e.g., $z_{\rm conf} > 0.5$) successfully constrains these systematic offsets within a markedly narrower envelope. This effective suppression of localized bias is critical for ensuring that the reconstructed redshift distribution does not introduce spurious signals. Consequently, the $z_{\rm conf}$ mechanism establishes a high-integrity data foundation for precision cosmological probes—such as weak gravitational lensing and galaxy clustering—thereby systematically reducing the localized calibration risks in the redshift distribution. Although residual systematic offsets inevitably persist in the point estimates, suppressing these localized fluctuations provides a cleaner sample subset, which may help to mitigate the propagation of redshift uncertainties into downstream cosmological analyses.


\section{Summary and Conclusions} \label{sec:conclusions}

Motivated by the rigorous requirements for high-precision photometric redshifts ($z_{\rm phot}$) and robust uncertainty quantification in next-generation surveys such as LSST, this study evaluates the LSTM-MDNz hybrid architecture using the HSC GalaxiesML dataset—a small-scale proxy for LSST-like surveys. By restructuring multi-band photometric data into wavelength-ordered sequences to emulate the physical continuity of SEDs and leveraging MDNs for explicit posterior PDF modeling, we demonstrate the efficacy of this architecture in resolving complex color–redshift degeneracies.

In terms of point-estimation accuracy and robustness, compared with the BNN baseline established by \citet{2024ApJ...964..130J}, our method achieves significant improvements across all key performance indicators. Specifically, the RMSE and MAE are reduced by approximately 10\% and 13\%, respectively; the normalized median absolute deviation ($\sigma_{\rm NMAD}$) decreases from 0.026 to 0.024. Notably, the outlier fraction ($O$) and the catastrophic outlier rate ($O_C$) are reduced by approximately 18\% and 22\%, respectively. These results indicate that when multi-band data are treated as a wavelength-ordered sequence, the LSTM gating mechanism more effectively captures inherent nonlinear correlations. By leveraging the physical continuity of SEDs, this sequential processing provides more consistent structural constraints than traditional fully connected architectures, leading to a more stable mapping in high-dimensional color space.

Regarding the uncertainty quantification and probabilistic calibration of photometric redshifts, the LSTM-MDNz model produces PDFs that demonstrate both robust and well-calibrated performance. The PIT distribution on the test set closely approximates a theoretical uniform distribution, confirming that the model’s predictive uncertainty estimates are statistically consistent and free from significant over-confidence or under-confidence biases. Concurrently, the median CRPS—a core metric for assessing the precision and sharpness of the predicted PDFs—remains stable at a low value of $0.0160$, indicating that the model achieves high predictive accuracy while maintaining physically plausible distribution widths. 

Unlike the unimodal Gaussian assumptions commonly employed in standard BNNs, the MDN backend in our model uses a GMM to explicitly characterize the multimodal degeneracies inherent in redshift space. This architecture provides the flexibility to capture non‑Gaussian posterior distributions arising from color–redshift degeneracies, effectively identifying and representing potential alternative redshift solutions. Through this precise modeling of complex probabilistic morphologies, LSTM-MDNz not only enhances the robustness of point estimates but also provides a more reliable statistical foundation for downstream cosmological probes—such as dark energy surveys—that are exquisitely sensitive to the accuracy of the $N(z)$ distribution.

Furthermore, the PDF-based confidence metric, $z_{\rm conf}$, introduced in this study offers an effective screening scheme for constructing high-purity redshift catalogs. Our analysis indicates that by rejecting a relatively small fraction (approximately 4\%–15\%) of low-confidence samples, the outlier fraction ($O$) can be drastically reduced from 6.5\% to below 1.3\%. This purification mechanism demonstrates enhanced efficiency compared to conventional strategies based solely on the width of unimodal distributions (e.g., $\sigma_z$) at equivalent sample retention levels. When evaluated across sequential $z_{\rm spec}$ intervals, this filtering approach demonstrates high robustness in suppressing systematic biases, narrowing the scatter, and limiting the accumulation of outliers at high redshifts ($z_{\rm spec} \gtrsim 0.8$), where color–redshift degeneracies tend to be more pronounced.

To decode the underlying drivers of these global performance gains, we have decoupled the relative contributions of the sequential network topology from the input features via a dedicated ablation analysis. Unlike traditional MLP-based frameworks, which can face geometric alignment limitations when attempting to form structured, localized associations between individual band magnitudes and their corresponding uncertainties, our sequential architecture seamlessly accommodates heteroscedastic noise by constructing the input as a wavelength-ordered, two-dimensional sequence vector, $\mathbf{x}_\lambda = [{\rm mag}_\lambda, \Delta{\rm mag}_\lambda]$. This design allows the LSTM units to dynamically weight each band based on its statistical signal-to-noise ratio (${\rm SNR}$), demonstrating that this uncertainty-aware capability is an inherent, structural characteristic of the architecture rather than a mere standalone data dividend.

This theoretical expectation is fully borne out by the performance of our magnitude-only baseline configuration, LSTM-MDNz (Mags Only). Even in the complete absence of cmodel magnitude uncertainty features, this pure-architecture variant still yields point estimates that remain statistically consistent with, or modestly superior to, the reference BNN framework---maintaining a comparable robust scatter and a stable outlier fraction while generating noticeably better-calibrated PIT distributions. This confirms the exceptional capacity of sequential encoding in capturing underlying physical SED gradients and highlights the advantage of the MDN backend in mitigating multi-modal degeneracies in color-redshift space. Concurrently, reintroducing these uncertainty features into the full model yields a clear synergistic effect: by providing adaptive probabilistic constraints for lower-${\rm SNR}$ sources, it systematically suppresses extreme estimation failures and drives down the global outlier population. Therefore, the overall success of the LSTM-MDNz framework can be attributed to the complementary interplay between its sequential modeling capability and its uncertainty-feature fusion mechanism.

While the LSTM-MDNz framework demonstrates reliable capabilities in sequential modeling and feature fusion, certain systematic boundaries must be carefully considered when extending its application to full LSST cosmological samples. Supervised machine learning estimators are intrinsically sensitive to the representativeness of their training data; existing spectroscopic reference catalogs are heavily affected by selection functions that favor brighter, higher-${\rm S/N}$ galaxies, whereas the core LSST cosmological analysis samples will inevitably include a vast population of fainter, higher-redshift sources. When encountering target populations whose underlying feature distributions deviate significantly from the training domain, sequence-based networks can experience a localized reduction in point-estimation accuracy or produce overly confident, miscalibrated PDF widths, as they attempt to extrapolate from unrepresentative color‑magnitude spaces. 

To facilitate the robust deployment of this framework within precision cosmology pipelines, future implementations will require integration with complementary mitigating methodologies. At the algorithmic level, implementing domain adaptation techniques—such as domain-adversarial neural networks—can encourage the extraction of domain-invariant features across varying depth scales, and photometric-space re-weighting via self-organizing maps (SOMs) can help adjust the empirical training weights to better align with the target wide-field survey distribution. Furthermore, at the cosmological pipeline level, incorporating these error-aware redshift representations into broader global frameworks, such as clustering-based redshift cross-calibrations, offers a promising avenue to safeguard against lingering dataset shifts and ultimately support the stringent statistical accuracy required for next-generation cosmic shear and clustering constraints.

In summary, the LSTM-MDNz framework, by combining sequential modeling with mixture density estimation, enables high-precision probabilistic modeling of galaxy photometric redshifts. As upcoming wide-field survey missions such as LSST and Euclid enter the era of "precision cosmology," the higher statistical fidelity and outlier resistance provided by this hybrid deep learning architecture will play a key role in constraining the nature of dark energy and the large-scale structure of the universe.

\begin{acknowledgements}
      Z.L. acknowledges the support from the National Natural Science Foundation of China (Grant No. 12573009) and the scientific research grants from the China Manned Space Project with Grand No. CMS-CSST-2025-A07 and CMS-CSST-2025-A05. H.X. acknowledges the support from the National Natural Science Foundation of China (NSFC 12203034), the Shanghai Science and Technology Fund (22YF1431500), the science research grants from the China Manned Space Project (CMS-CSST-2025-A07), and the Shanghai Municipal Education Commission regarding artificial intelligence empowered research. S.Z. acknowledges support from the National Natural Science Foundation of China (Grant No. 12173026), the Program for Professor of Special Appointment (Eastern Scholar) at Shanghai Institutions of Higher Learning, and the Shuguang Program (23SG39) of the Shanghai Education Development Foundation and Shanghai Municipal Education Commission. This work is also supported by the National Natural Science Foundation of China under Grant No. 12141302. 
\end{acknowledgements}

\bibliographystyle{aa}  
\bibliography{ref} 

%

\end{document}